\documentclass[conference]{IEEEtran}
\IEEEoverridecommandlockouts
\usepackage{cite}
\usepackage{amsmath,amssymb,amsfonts}
\usepackage{algorithmic}
\usepackage{graphicx}
\usepackage{textcomp}
\usepackage{xcolor}
\def\BibTeX{{\rm B\kern-.05em{\sc i\kern-.025em b}\kern-.08em
		T\kern-.1667em\lower.7ex\hbox{E}\kern-.125emX}}
\usepackage{amsthm}

\newtheorem{theorem}{Theorem}
\usepackage{graphicx}

\usepackage{color}
\usepackage[latin1]{inputenc}
\usepackage{multirow}
\usepackage{graphicx}
\usepackage{epstopdf}
\usepackage{bbm}     
\usepackage{amsmath, amssymb,amstext}
\usepackage{bm}
\usepackage{amsmath}
\usepackage{amsfonts}
\usepackage{booktabs} 
\usepackage{threeparttable} 
\usepackage{tabularx}
\usepackage{subfigure}
\usepackage{float}
\usepackage{url}
\usepackage{cuted}
\usepackage[ruled,linesnumbered,titlenumbered]{algorithm2e}
\usepackage[linkcolor=black,citecolor=black,urlcolor=black,colorlinks=true]{hyperref}

\title{A Guidance and Maneuvering Control System Design with Anti-collision Using Stream Functions with Vortex Flows for Autonomous Marine Vessels}
\author{Hongyu Zhou, Zhengru Ren, Mathias Marley, and Roger Skjetne
	\thanks{The work is partly sponsored by the Research Council of Norway (RCN) through the Centre for Research-based Innovation scheme, SFI AutoShip (RCN project 309230) and SFI MOVE (RCN project 237929), and the Centre of Excellence on Autonomous Marine Operations and Systems (NTNU AMOS, RCN project 223254). \textit{(Corresponding author: Zhengru Ren)}}
	\thanks{H. Zhou is with the Department of Aerospace Engineering, University of Michigan, Ann Arbor, MI 48109-2140, and also with the Department of Marine Technology, Norwegian University of Science and Technology (NTNU), NO-7491 Trondheim, Norway. (e-mail: zhouhy@umich.edu).}
	\thanks{Z. Ren is with the Institute for Ocean Engineering, Shenzhen International Graduate School, Tsinghua University, Tsinghua Campus, University Town, Shenzhen 518055, China, and also with the Centre for Research-based Innovation on Marine Operations (SFI MOVE), Department of Marine Technology, NTNU, NO-7491 Trondheim, Norway.  (e-mail: zhengru.ren@sz.tsinghua.edu.cn). }
	\thanks{M. Marley is with the Centre for Autonomous Marine Operations and Systems (AMOS), Department of Marine Technology, NTNU, NO-7491 Trondheim, Norway (e-mail: mathias.marley@ntnu.no).}
	\thanks{R. Skjetne is with the SFI MOVE, AMOS, and SFI AutoShip, Department of Marine Technology, NTNU, NO-7491 Trondheim, Norway (e-mail: roger.skjetne@ntnu.no).}}

\begin{document}

\maketitle
\thispagestyle{empty}
\pagestyle{empty}

\begin{abstract}
	
		Autonomous marine vessels are expected to avoid inter-vessel collisions and comply with the international regulations for safe voyages. This paper presents a stepwise path planning method using stream functions. The dynamic flow of fluids is used as a guidance model, where the collision avoidance in static environments is achieved by applying the circular theorem in the sink flow. We extend this method to dynamic environments by adding vortex flows in the flow field. The stream function is recursively updated to enable ``on the fly" waypoint decisions. The vessel avoids collisions and also complies with several rules of the Convention on the \textit{International Regulations for Preventing Collisions at Sea}. The method is conceptually and computationally simple and convenient to tune, and yet versatile to handle complex and dense marine traffic with multiple dynamic obstacles. The ship dynamics are taken into account, by using B\'{e}zier curves to generate a sufficiently smooth path with feasible curvature. Numerical simulations are conducted to verify the proposed method.
		
\end{abstract}


\section{Introduction}
In autonomous robotics, path planning is a central problem. The objective of path planning is often to find a safe path from the agent's initial position to a destination. In the marine industry, research on autonomous ships is gaining increasing attention, as a next evolutionary step in ship technology and operations. Conflict resolution, including how to find a collision-free path, is a main issue to be addressed \cite{huang2020ship} to achieve safe ship navigation and maneuvering. To find a path to guide the vessel to its destinations dynamically, the path-planning method should be able to characterize and update the surrounding terrain, obstacles, and any other relevant exterior information. A model incorporating such information, on which a path or maneuver can be planned, is called a \emph{guidance model} hereafter.

An abundance of path-planning methods have been reviewed in \cite{huang2019collision,patle2019review, mac2016heuristic}, including the search-based and sampling-based methods. The artificial potential field (APF) method~\cite{khatib1986real} is another classical guidance model for navigation, widely applied for real-time collision-free path planning \cite{ge2002dynamic,zhao2006robot,yin2008improved,huang2009velocity,ren2022dynamic,tan2010navigation,chiang2015path,chen2016uav,lyu2017ship}. In this method, the workspace of the vessel is represented by an artificial potential field where the destination attracts the vessel and the obstacles repulse the vessel. An advantage of the APF method is that it can easily be modified to incorporate more information, such as velocity \cite{ge2002dynamic}, acceleration \cite{yin2008improved}, and navigation rules \cite{lyu2017ship}. However, a vessel can be trapped in a local extremum and not be able to reach its destination.

However, the path found by the abovementioned methods might not be suitable for marine vessels. A sufficiently smooth path is desired, typically requiring that the three first derivatives are continuous and bounded \cite{skjetne2005maneuvering}. The search-based and sampling-based methods can highly depend on
the resolution of the discretization and can easily result in non-smooth trajectories, possibly generating infeasible or impractical direction changes. Moreover, a set of preset rules, such as the \textit{International Regulations for Preventing Collisions at Sea} (COLREGs) \cite{COLREGs}, should be considered to guide path planning and collision avoidance.

The stream function, also called the potential flow, is a model derived from hydrodynamics and can be applied in path planning \cite{milne1996theoretical,axler2013harmonic,currie2016fundamental}. Stream functions for path planning are a subset of the APF methods, since they generate a vector field for the vessel to follow. While unlike the APF methods using attractive and repulsive forces for navigation, stream functions do not suffer from the problem of local extremum. The basic idea of this method is to model a stream of fluid flowing around circular obstacles to obtain the trajectory of fluid particles, called streamlines. Since streamlines are collision-free paths, modeling the flow based on potential flow theory enables one to find a safe and smooth path \cite{waydo2003vehicle,ye2005coordinated,hu2007autonomous,pedersen2012marine}. This method is computationally efficient, and yet it extends well to complex situations by applying potential flow theory. Existing works focus on static obstacles; see \cite{kim2011stream,pedersen2012marine}. The collision avoidance solution in existing works is to simply follow the streamlines; however, these can be dynamically infeasible for marine vessels. Besides, there are few attempts to apply the stream function in dynamically changing environments. To the best of our knowledge, collision avoidance with moving obstacles using stream function is only investigated in \cite{waydo2003vehicle}, while the method is only demonstrated in the case with one moving obstacle.

In recent years, some existing navigation rules have been used to find a rule-compliant collision-free solution; for example  \cite{perera2012intelligent,kuwata2013safe,praczyk2015neural,johansen2016ship,lyu2017ship,lyu2019colregs}. Among these, \cite{lyu2017ship} and \cite{lyu2019colregs} use APF and suffer from local minima. Additionally, the dynamics of ships are assumed to be holonomic. \cite{kuwata2013safe} applies the ``velocity obstacles'' method, where irregular shape of the obstacles, sensing errors, and the COLREGs are taken into account. However, it also neglects the dynamics and kinetics constraints of ships. In \cite{perera2012intelligent} and \cite{praczyk2015neural}, rule-compliant actions are generated by neural networks. The limitation is that collision-free solutions are not guaranteed. If some situations are not included in the training database, the generated actions can cause collision. Using simplified model predictive control (MPC), \cite{johansen2016ship} manages to handle complex scenarios with multiple randomly moving obstacles. However, the MPC structure usually increases computation time, which can be a problem for real-time applications.

In this paper, we propose a stepwise (recursive) path planning method based on a guidance model derived from the stream function. Waypoints are determined in a stepwise manner, using the recursively updated stream function incorporating the latest surrounding traffic and workspace information. Unlike the existing work, we use stream functions as an intermediate decision model for waypoint selection and do not force the vessel to follow the streamline directly. By adding a component called vortex flow \cite{milne1996theoretical,axler2013harmonic,currie2016fundamental}, we demonstrate that the proposed method can be applied in dynamic environments with multiple moving obstacles. 
To illustrate how the path-planning algorithm works, a path is generated from the waypoints using B\'{e}zier curves, and a path-following controller is designed to make the ship move along the generated path. The environmental loads are not considered \cite{ren:20216DOFIdentification}. The contributions of this paper are summarized as follows:

\begin{itemize}
	\item We extended the stream function approach by introducing vortices, where the flow field around the obstacle ships are modified by vortex flows. Together with the proposed optimal waypoint selection approach, the method ensures compliance with COLREGs rules in complex marine traffic. This cannot be achieved by using typical methods based on stream functions.
	
	\item The stepwise path planning is integrated into the closed-loop guidance and control system, where the vessel dynamics is considered in the control design.
	
	\item The performance of the proposed method in various complex and dense marine traffic is verified through simulations.
\end{itemize}

The rest of the paper is organized as follows. Section 2 formulates the collision-avoidance problem. Section 3 describes the stream function method for waypoint generation. Section 4 presents the path-generation method. Section 5 presents the simulation results. Section 6 concludes the paper with directions for future work.

\textbf{Notation}: The sets of real numbers and real positive numbers are $\mathbb{R}$ and $\mathbb{R}^+$, respectively. A curve with continuous and bounded $i^{th}$ derivatives is denoted by $\mathcal{C}^i$. The Euclidean vector norm is $|x| := (x^\top x)^{1/2}$. Total time derivatives of $x(t)$ up to order $n$ are denoted $\dot{x}$, $\ddot{x}$, ${x}^{(3)}$, $\ldots$, ${x}^{(n)}$.  Partial differentiation is denoted by a superscript: $\alpha^t (x, \theta, t) := \partial \alpha / \partial t$, $\alpha^{x^2} (x, \theta, t) := \partial^2 \alpha / \partial x^2$,  $\alpha^{\theta^n} (x, \theta, t) := \partial^n \alpha / \partial \theta^n$, etc.

\section{Problem Formulation}
\label{Ch2}

\subsection{System description}
\begin{figure}[h]
	\centering
	\includegraphics[width=0.8\linewidth]{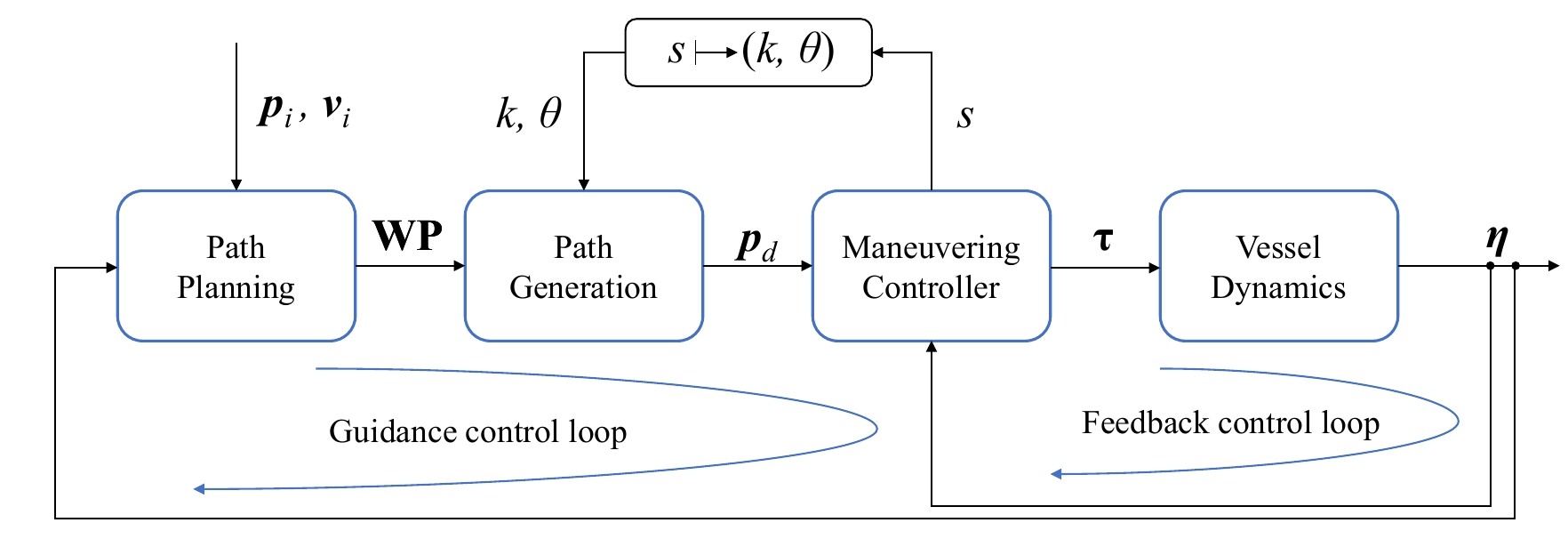}
	\caption{The system architecture for a marine craft.\label{GNC}}
\end{figure}

The proposed system in this paper is illustrated in Figure \ref{GNC}. The path planning module determines discrete waypoints $\mathbf{WP}$, among a set of candidate points, to be used by the path generation module to generate the desired path $\mathbf{b}: [0,1] \rightarrow \mathbb{R}^2$ with path parametrization variable $\theta$ to determine the desired position $\bm{p}_d(\theta) \in \mathbb{R}^2$ along the path segment $k$. The maneuvering controller~\cite{skjetne2005maneuvering} calculates the desired forces and moments $\bm\tau\in \mathbb{R}^3$ to track the desired path and path speed. 

There are several obstacles in the workspace. We assume each obstacle moves with a constant speed and heading (i.e., not following COLREGs), and the positions and velocities of the vessel, destination, and obstacles are known. The proposed control system aims to guide and control the vessel to the destination with collision avoidance and compliance to regulations. Three rules from COLREGs \cite{COLREGs} are selected to assess the proposed system:

\begin{itemize}
	
	\item Rule 13 $-$ Overtaking: \textit{Any vessel overtaking any other shall keep out of the way of the vessel being overtaken.}
	
	\item Rule 14 $-$ Head-on situation: \textit{When two power-driven vessels are meeting on reciprocal or nearly reciprocal courses so as to involve risk of collision, each shall alter her course to starboard so that each shall pass on the port side of the other.} 
	
	\item Rule 15 $-$ Crossing situation: \textit{When two power-driven vessels are crossing so as to involve risk of collision, the vessel which has the other on her own starboard side shall keep out of the way and shall, if the circumstances of the case admit, avoid crossing ahead of the other vessel.} 
	
\end{itemize}

\subsection{Workspace representation}
\begin{figure}
	\centering
	\includegraphics[width=0.8\linewidth,height=0.7\linewidth]{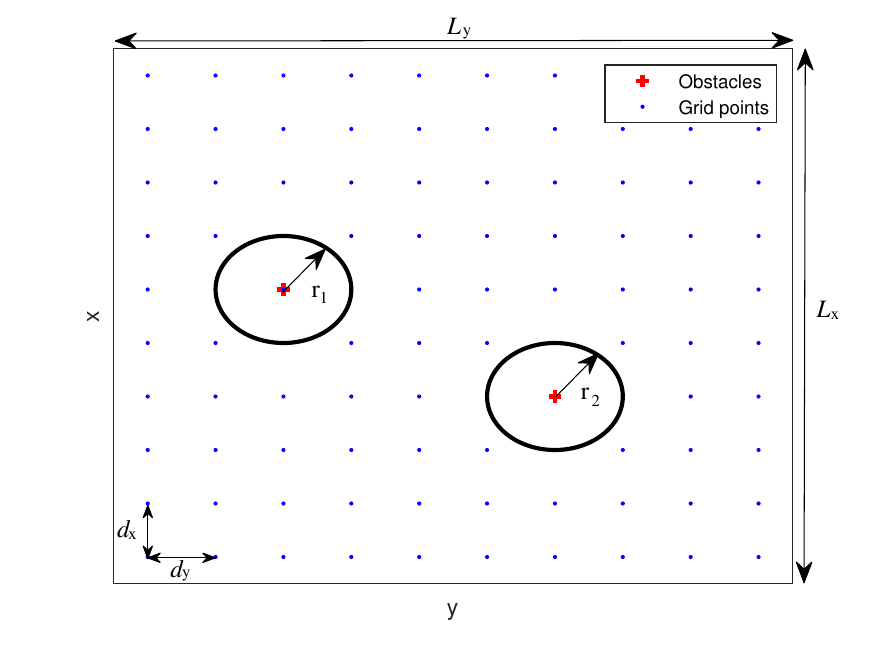}
	\caption{Illustration of a workspace representation.}
	\label{fig_Workspace}
\end{figure}

The workspace is represented by a discrete grid in the North-East (NE) reference frame. An example is shown in Figure \ref{fig_Workspace}. The size of the workspace is $L_x(m) \cdot L_y(m)$. We place $N_x$ equidistant horizontal rows with an interval of $d_x = \frac{L_x}{N_x}$ in the $x$-direction. Each row has $N_y$ equidistant discrete points with an interval of $d_y = \frac{L_y}{N_y}$ in the $y$-direction. This results in $N = N_x N_y$ grid points in total. We define the set of grid point as
\begin{equation}
	\mathcal{P}_g := \left\{(x_{m,n},y_{m,n}) \in \mathbb{R}^2:  (m,n)\in \mathcal{N}_x \times \mathcal{N}_y \right\},
	\label{eq_grid_point_set}
\end{equation}
where $\mathcal{N}_x=\{1,2,...,N_x\}$ and $\mathcal{N}_y=\{1,2,...,N_y\}$ are index sets.

There is a number $N_o$ of obstacles, and only one destination in the workspace. Each obstacle is represented by a circular domain with radius $r_i$, where $i\in \mathcal{N}_o := \{1,2,...,N_o\}$. We focus on circular obstacles in this paper. Note that any star-shaped region can be diffeomorphically transformed into a sphere \cite{rimon1990exact, rimon1991construction}. It is assumed that all obstacles are distributed without overlapping, and that the positions and velocities of the obstacles are known.

We will later refer to our vessel as `own ship' (OS). If an obstacle is a COLREG-compliant motor-powered ship, we will refer to this as a `target ship' (TS); otherwise, it is only referred to as an obstacle.

\subsection{Problem statement}
Given the positions of the destination and obstacles, the path planning module aims to find a set of discrete waypoints to guide the vessel towards the destination while avoiding collision. The path should comply with rules 13-15 from COLREGs. Generally, the position information of the destination and obstacles is incorporated into the path planning method.

In the workspace information, it is assumed that the known locations of the destination and obstacles are denoted by $\bm{p}_t = [x_t \  y_t]^\top \in~\mathcal{P}_g$ and  $\bm{p}_i = [x_{i}  \  y_{i}]^\top$, respectively. The corresponding velocity of the $i^{th}$ obstacle is $\bm{v}_i = [x_{v_i}  \  y_{v_i}]^\top$, $i\in \mathcal{N}_o $. The position of the current waypoint $\mathbf{WP}_k=[x_k \  y_k]^\top\in\mathcal{P}_g$, where $k$ is the index of the waypoint. The path planner module is designed to find the next waypoint $\mathbf{WP}_{k+1}=[x_{k+1} \  y_{k+1}]^\top\in~\mathcal{P}_g$ to generate a new path segment with sufficient continuity at the waypoints. Since the obstacles are usually moving, the path planner module should update the workspace information and find the next waypoint using the latest information. Once the vessel approaches the next waypoint, the path planner module can use the latest workspace information to determine the new waypoint. This procedure shall be done recursively until the vessel reaches the destination.

\section{Stream function guidance model}
The proposed guidance model for path-planning is inspired by the stream function, which is widely applied in hydrodynamics. In this section, some important concepts from hydrodynamics are introduced \cite{milne1996theoretical,axler2013harmonic,currie2016fundamental}. Based on these concepts, the workspace is characterized and the algorithm to recursively generate the waypoints using a stream function is presented. 

\subsection{Potential flows and complex potential}
A 3D flow is irrotational if the vorticity vector ${\omega}$ is zero everywhere in the fluid, which is given by
\begin{equation}
	{\omega} = \nabla \times \nu_f = \left[ \frac{\partial}{\partial x} \    \frac{\partial}{\partial y} \   \frac{\partial}{\partial z}\right]^\top \times [\nu_x \  \nu_y \ \nu_z]^\top = 0,
	\label{eq_irrotational}
\end{equation}
where $\nabla$ denotes the gradient with respect to Cartesian coordinates, $\times$ means cross product, and $\nu_f = [\nu_x \  \nu_y \ \nu_z]^\top$ is the velocity vector of the flow. In a 2D flow where $\nu_z = 0$, eq. \eqref{eq_irrotational} is simplified to be 
\begin{equation}
	{\omega} =  \frac{\partial \nu_y}{\partial x}- \frac{\partial \nu_x}{\partial y} = 0.
	\label{eq_irrotational_2D}
\end{equation}

For any scalar velocity potential function $\phi_f$, we have
\begin{equation}
	\nabla \times \nabla \phi_f = 0.
	\label{eq_scalar}
\end{equation}
The irrotational flow fields are called the potential flows, represented by $\nu_f = \nabla \phi_f$. The velocity of the flow can be calculated from a velocity potential, 
\begin{equation}
	\begin{aligned}
		\nu_x = \frac{\partial \phi_f}{\partial x}, \quad
		\nu_y = \frac{\partial \phi_f}{\partial y}.
	\end{aligned}
	\label{eq_velocity}
\end{equation}

The velocity potential satisfies Laplace's equation, $\nabla^2 \phi_f = 0$, since an ideal flow must satisfy the condition of continuity in Cartesian coordinates, i.e., $\nabla \nu_f = 0$. This condition is expressed as 
\begin{equation}
	\frac{\partial \nu_x}{\partial x}+\frac{\partial \nu_y}{\partial y} = 0.
\end{equation}

A stream function $\psi_f$ is defined such that
\begin{equation}
	\begin{aligned}
		\nu_x = \frac{\partial \psi_f}{\partial y}, \quad
		\nu_y = -\frac{\partial \psi_f}{\partial x}.
	\end{aligned}
	\label{eq_stream} 
\end{equation} 

Then a stream function also satisfies Laplace's equation by definition, i.e., $\nabla^2 \psi_f = 0$. A streamline is the line along a constant value of $\psi_f$. Physically, a streamline is the trajectory of a fluid particle as it moves in the flow. By combining \eqref{eq_velocity} and \eqref{eq_stream}, the velocity potential and the stream function satisfy the Cauchy-Riemann equation \cite{riemann1867grundlagen}, i.e.,
\begin{equation}
	\begin{aligned}
		\frac{\partial \phi_f}{\partial x}= \frac{\partial \psi_f}{\partial y}, \quad
		\frac{\partial \phi_f}{\partial y} = -\frac{\partial \psi_f}{\partial x}.
	\end{aligned}
\end{equation} 

\begin{figure}
	\centering
	\includegraphics[width=0.7\linewidth,height=0.63\linewidth]{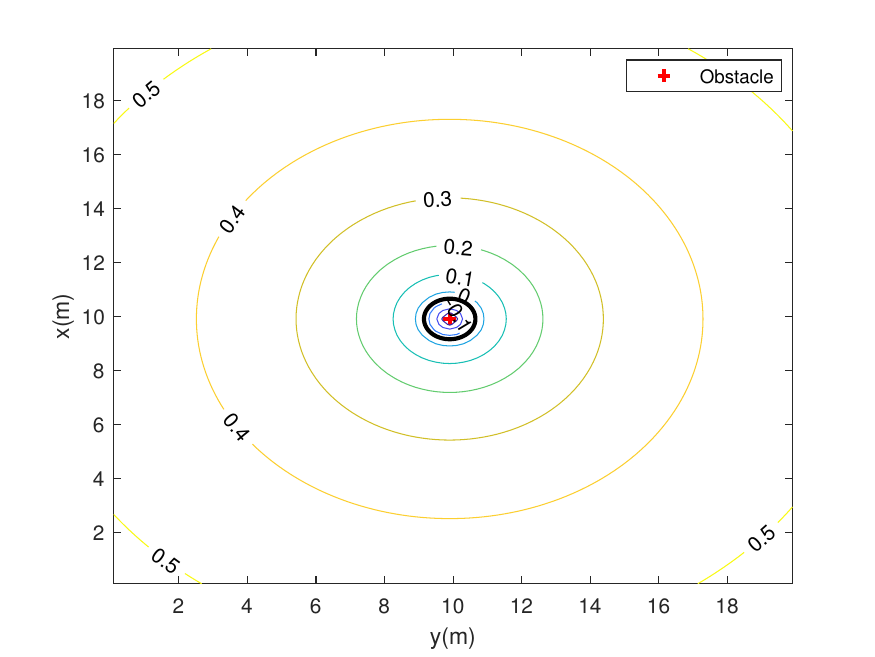}
	\caption{The streamlines of a vortex flow with $C>0$.}
	\label{Ex_vortex}
\end{figure}

With the velocity potential and the stream function, we can now define an irrotational and inviscid 2D flow $\omega_f(\mathcal{Z})$ as
\begin{equation}
	\omega_f(\mathcal{Z}) = \phi_f + \mathrm{i}\psi_f,
\end{equation} 
where $\mathcal{Z} = x+\mathrm{i}y$ with $\mathrm{i}$ as the imaginary unit. We can see that the real part of $\omega_f(\mathcal{Z})$ is the velocity potential and the imaginary part is the stream function. The uniform flow, sink, source, and vortex are the most important flow types.
\begin{itemize}
	\item For a uniform flow in the $x$-direction with $\nu_f = [\nu_x \  0 \ 0]^\top$, the complex potential is given by
	\begin{equation}
		\omega_f(\mathcal{Z}) = \nu_x  \mathcal{Z} = \nu_x x + \mathrm{i} \nu_x y.
		\label{uni_flow} 
	\end{equation} 
	
	\item The complex potentials for sources and sinks are 
	\begin{equation}
		\omega_f(\mathcal{Z}) = C \ln(\mathcal{Z}) = \frac{C}{2}\ln{(x^2+y^2)}+\mathrm{i}C\text{atan}\left(\frac{y}{x}\right),
		\label{sink_source} 
	\end{equation} 
	where $C>0$ or $C<0$ denote source or sink, respectively. 
	
	\item The complex potential for a vortex is 
	\begin{equation}
		\omega_f(\mathcal{Z}) = \mathrm{i}C \ln(\mathcal{Z}) = C\text{atan}\left(\frac{y}{x}\right)-\mathrm{i}\frac{C}{2}\ln{(x^2+y^2)},
		\label{vortex_flow} 
	\end{equation} 
	where $C>0$ and $C<0$ indicate the fluids rotating around the vortex center in counterclockwise and clockwise directions, respectively. An example is given in Figure \ref{Ex_vortex}, where the numbers on the streamlines indicate the corresponding values of the stream functions. Note that the value along the boundary of a circular obstacle is constant.
\end{itemize}

\subsection{Obstacle representation}
\subsubsection{Circle theorem}
The circle theorem is adopted to represent a circular obstacle in the flow. The theorem is valid for an obstacle at an arbitrary position. By applying the theorem, the boundary condition on the obstacle is satisfied, i.e., $\psi_f$ is constant along the boundary of the obstacle. 

\begin{theorem}[Circle theorem \cite{milne1996theoretical}]
	Let there be an irrotational two-dimensional flow of incompressible inviscid fluid in the $\mathcal{Z}$-plane. Let there be no rigid boundaries and let the complex potential of the flow be $f(\mathcal{Z})$, where the singularities of $f(\mathcal{Z})$ are all at a distance greater than $\mathrm{a}$ from the point $\mathrm{b}$ in the $\mathcal{Z}$-plane. Let $\Bar{f}(\mathcal{Z})$ and $\Bar{\mathrm{b}}$ be the conjugate function of $f(\mathcal{Z})$ and $\mathrm{b}$, respectively. If a circular cylinder, typified by its cross-section $|\mathcal{Z}-\mathrm{b}| = \mathrm{a}$, is introduced into the flow, the complex potential becomes
	\begin{equation}
		\omega_f(\mathcal{Z}) = \phi_f + \mathrm{i}\psi_f = f(\mathcal{Z}) + \Bar{f}\left(\frac{\mathrm{a}^2}{\mathcal{Z}-\mathrm{b}}+\Bar{\mathrm{b}}\right).
		\label{eq_Circletheorem} 
	\end{equation} 
\end{theorem}

By applying the circular theorem, we can obtain the stream function with one obstacle in a basic flow. For example, using equations \eqref{uni_flow} and \eqref{eq_Circletheorem}, the stream function for a circular obstacle centered at the origin in a uniform flow is given by 
\begin{eqnarray}
	\psi_{f} = Cy\left(1-\frac{r_i^2}{x^2+y^2}\right).
	\label{SF_uniform_eq}
\end{eqnarray}

Using equations \eqref{sink_source} and \eqref{eq_Circletheorem}, the circular obstacle in a sink flow is given by
\begin{eqnarray}
	\psi_{f} = -C \text{atan} \left(\frac{y}{x} \right)+C \text{atan} \left( \frac{\frac{r_i^2(y-y_{i})}{(x-x_{i})^2+(y-y_{i})^2}+y_{i}}{\frac{r_i^2(x-x_{i})}{(x-x_{i})^2+(y-y_{i})^2}+x_{i}}\right),
	\label{SF_sink_eq}
\end{eqnarray}
where $(x_{i},y_{i})$ is the position of the $i$th obstacle. Equation \eqref{SF_sink_eq} is often used for path planning when the destination is modeled as a sink flow.

\subsubsection{Multiple obstacles}
In the flow field, if there are multiple circular obstacles, the Laplace's equation with multiple boundary conditions must be solved to obtain the stream function. However, this is analytically impossible \cite{waydo2003vehicle}. In the case of multiple obstacles, a method called \textit{addition and thresholding} is applied \cite{waydo2003vehicle,sullivan2003using}.

The basic idea of \textit{addition and thresholding} is that each obstacle has its influence area. The stream function at a grid point $\bm{p}_g = [x_g\ y_g]^\top \in~\mathcal{P}_g$ is influenced by the obstacles whose distance to $\bm{p}_g$ is less than or equal to a range $l_{i}$, i.e., $\left| \bm{p}_g-\bm{p}_i \right| \leq l_{i}$, $i\in \mathcal{N}_o$, with $l_{i} \geq r_{i}$. An example is shown in Figure \ref{Ex_moving_vortex_a}. 

To apply this method in an environment with $N_o$ obstacles, we assume that $\psi_{i}$ is the stream function generated by only considering the $i$th obstacle in the sink flow. The stream function of multiple fixed obstacles in the whole workspace is given by
\begin{eqnarray}
	\psi_f(\bm{p}_g)=  \begin{cases}
		&\sum_{i=1}^{N_o}  c_i\psi_{i}(\bm{p}_g),
		\\& \qquad \text{if } \exists i \in \mathcal{N}_o  \text{ s.t. } {\left| \bm{p}_g-\bm{p}_i \right| \leq l_{i}}, \\
		&\sum_{i=1}^{N_o}\psi_{i}(\bm{p}_g), \text{ otherwise},
	\end{cases}
\end{eqnarray}
where 
\begin{eqnarray}
	c_i = \left\{ \begin{array}{ccl}
		1, & & \text{if} \ {\left| \bm{p}_g-\bm{p}_i \right| \leq l_{i}}\\
		0, & & \text{if} \ {\left| \bm{p}_g-\bm{p}_i \right| > l_{i}}
	\end{array}. \right.
	\label{SF_ci}
\end{eqnarray}

This method removes the influence of those obstacles that should not have a direct impact on the streamlines, while the influences of the other obstacles are added together to create the vector field.

\subsubsection{Collision avoidance of multiple moving obstacles}
\begin{figure}[t]
	\centering
	\includegraphics[width=0.7\linewidth]{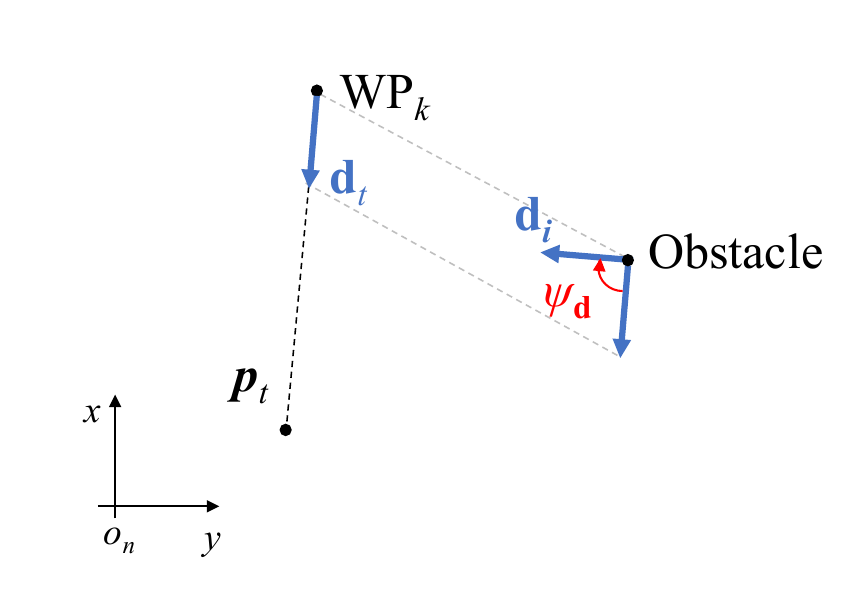}
	\caption{Main information used for calculating $S_i(\psi_{\mathbf{d}})$.}
	\label{vortex_dir}
\end{figure}

For the avoidance of multiple moving obstacles, additional vortex flows with centers at $\bm{p}_i$ are added to modify the original stream function. For all $\bm{p}_g$ such that $\left| \bm{p}_g-\bm{p}_i \right| = r_i$, the stream function of a vortex flow is constant, which implies that the boundary conditions on the obstacles can be satisfied.

The stream function of a vortex flow is 
\begin{eqnarray}
	\psi_{\bm{v}_i}(\bm{p}_g) = C_{\bm{v}_i} S_i(\psi_{\mathbf{d}})\left| \bm{v}_i \right|\ln(\left| \bm{p}_g-\bm{p}_i \right| ^2),
\end{eqnarray}
where $C_{\bm{v}_i} \in \mathbb{R}^+$ is a positive tuning parameter that combines with $\left| \bm{v}_i \right|$ represents the strength of the vortex flow, $S_i(\psi_{\mathbf{d}})$ is a signed function determining the direction of the vortex, given by

\begin{eqnarray}
	S_i(\psi_{\mathbf{d}}) = \left\{\begin{array}{cc}
		1, \quad \text{if} \ i^{th} \  \text{obstacle is COLREG-compliant} \\
		\left\{\begin{array}{cc}
			-1,  & \text{if} \  \psi_{\mathbf{d,l}} < \psi_{\mathbf{d}} < \psi_{\mathbf{d,u}}  \\
			1,  & \text{otherwise}
		\end{array}, \right.  \ \  \text{otherwise}.
	\end{array} \right.   
	\label{eq_Sd}
\end{eqnarray}
Here, $\psi_{\mathbf{d}} \in [-\pi, \pi)$ is the angle between the vectors $\mathbf{d}_i$ and $\mathbf{d}_t$, $\psi_{\mathbf{d,l}}$ and $\psi_{\mathbf{d,u}}$ (with $0 < \psi_{\mathbf{d,l}} < \psi_{\mathbf{d,u}} < \pi$) are tuning parameters, $\mathbf{d}_t=\frac{\bm{p}_t-\mathbf{WP}_k}{\left| \bm{p}_t-\mathbf{WP}_k \right|}$ is the unit vector pointing from the current waypoint $\mathbf{WP}_k$ to the destination, and $\mathbf{d}_i=\frac{\bm{v}_i}{\left| \bm{v}_i \right|}$ is the unit vector with the direction of the obstacle's velocity. This is illustrated in Figure~\ref{vortex_dir}.

The function $S_i(\psi_{\mathbf{d}})$ is used to assign a port or starboard turn to pass obstacles. Combined with a proper logic, this enables COLREGs compliance. When $\mathbf{d}_i$ nearly aligns with  $\mathbf{d}_t$, which indicates a head-on or overtaking situation, a counterclockwise vortex is added that forces the vessel (own ship; OS) to move towards its starboard. In a crossing situation, the obstacle might cross from either the vessel's starboard side or port side. When the obstacle crosses from the starboard side, $S_i(\psi_{\mathbf{d}})=1$ and a counterclockwise vortex is generated, so that the vessel can give way to the obstacle. If the obstacle crosses from the port side, which is classified as $\psi_{\mathbf{d}} \in (\frac{\pi}{4}, \frac{3\pi}{4})$ in Eq.~\eqref{eq_Sd}, $S_i(\psi_{\mathbf{d}})=-1$ and a clockwise vortex is generated. If the $i^{th}$ obstacle complies with COLREGs rules (e.g., the obstacle is a target ship), then $S_i(\psi_{\mathbf{d}})$ is set to be 1, without further modification, to make the vessel stand on its course with respect to the obstacle. More sophisticated logic for COLREGs compliance can be considered for implementation, using for instance the classification proposed in \cite{thyri2020reactive,eriksen2020hybrid}.

The resulting stream function for the whole workspace becomes
\begin{eqnarray}
	\psi_f(\bm{p}_g)=  \begin{cases}
		&\sum_{i=1}^{N_o}  c_i(\psi_{i}(\bm{p}_g)+\psi_{\bm{v}_i}(\bm{p}_g)),
		\\& \qquad \text{if } \exists i \in \mathcal{N}_o  \text{ s.t. } {\left| \bm{p}_g-\bm{p}_i \right| \leq l_{i}}, \\
		&\sum_{i=1}^{N_o}(\psi_{i}(\bm{p}_g)+\psi_{\bm{v}_i}(\bm{p}_g)), \text{ otherwise},
	\end{cases}
	\label{eq_stream_func_vortex}
\end{eqnarray}
where $c_i$ has the same definition as in Eq.~\eqref{SF_ci}.

\begin{figure}[htb!]
	\centering
	\subfigure[]{
		\includegraphics[width=0.7\linewidth]{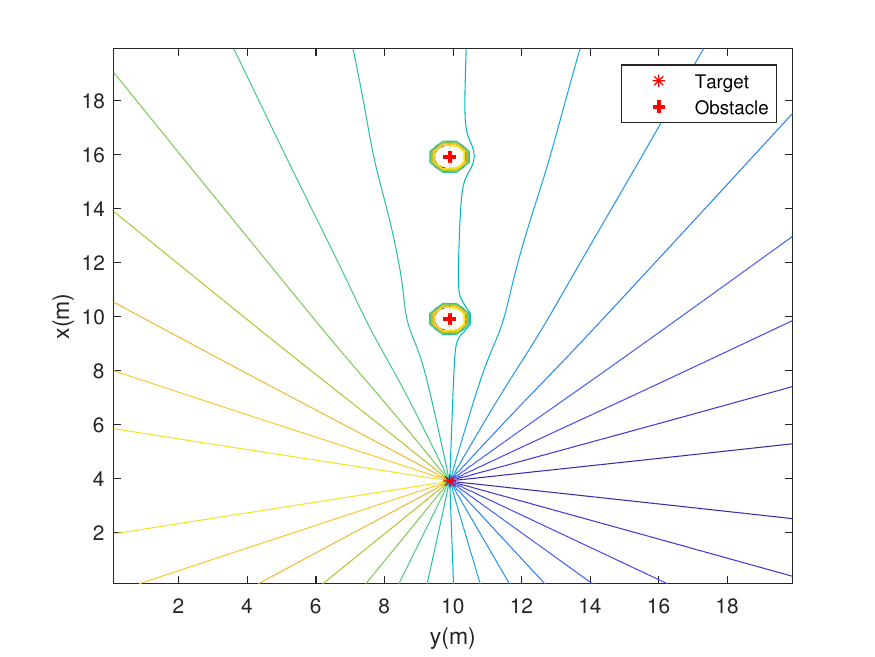}
		\label{Ex_moving_vortex_a}
	} \\ \vspace{-1em}
	\subfigure[]{
		\includegraphics[width=0.7\linewidth]{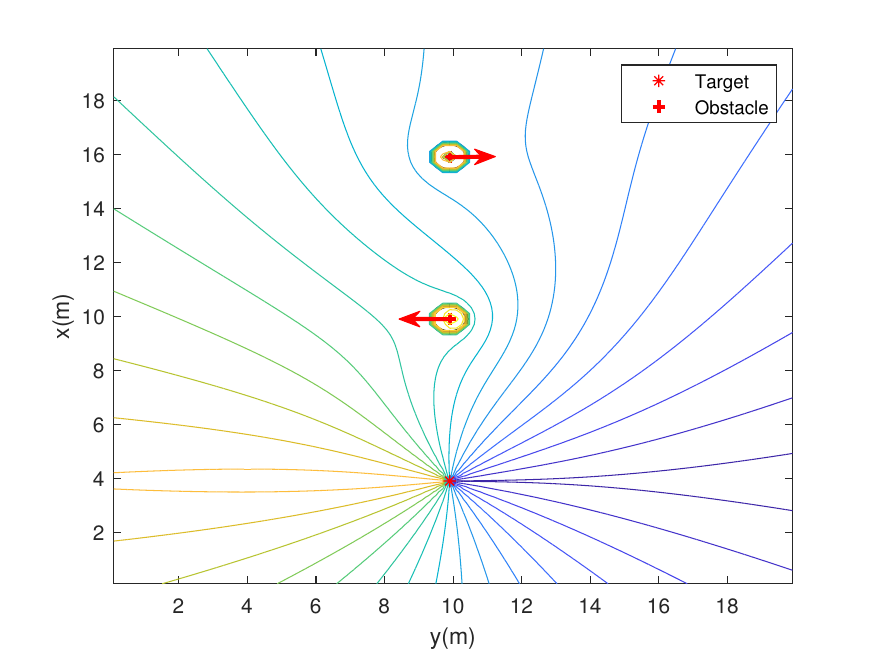}
		\label{Ex_moving_vortex_b}
	} \\ \vspace{-1em}
	\subfigure[]{
		\includegraphics[width=0.7\linewidth]{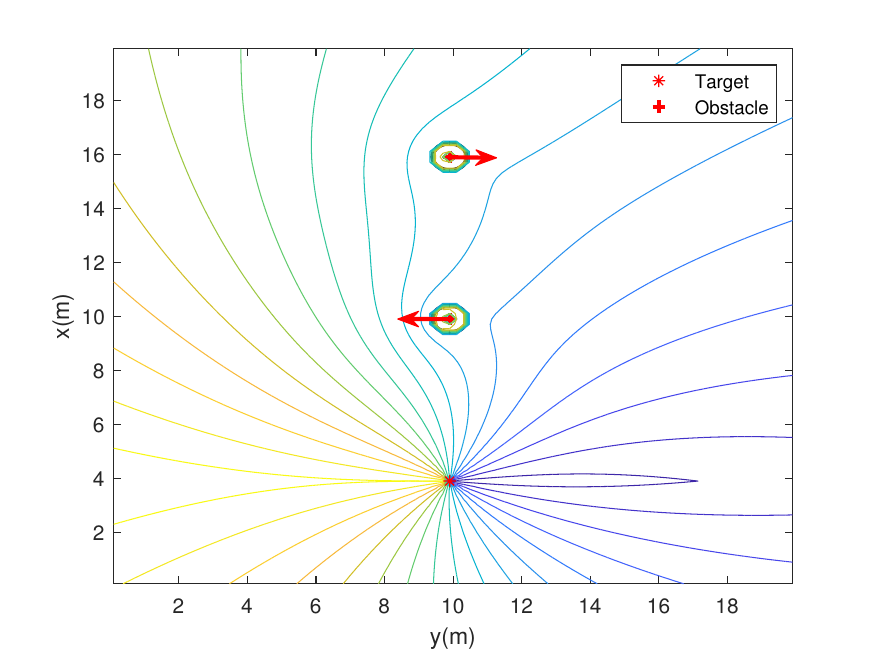}
		\label{Ex_moving_vortex_c}
	}
	\caption{Avoidance of multiple moving obstacles: The flow field (a) without vortex; (b) obstacle avoidance, with vortex strength $0.05$ (counterclockwise) and $-0.05$ (clockwise), and (c) COLREG-compliance, with vortex strength $0.05$ (counterclockwise) and $0.07$ (counterclockwise). The streamlines are calculated by using the  \textit{addition and thresholding} method. The red arrows indicate the directions of obstacles' velocities.}
	\label{Ex_moving_vortex} 
\end{figure}

By adding vortex flows with well-tuned $C_{\bm{v}_i}$, the streamlines close to the obstacles will be moved towards the opposite side of the direction of motion of the obstacle. A comparison between the flow field with and without adding the vortex flows is shown in Figure \ref{Ex_moving_vortex}. In this way, using a stream function updated recursively gives collision avoidance in dynamic environments. The streamlines far away from the obstacles are only slightly influenced by the vortex flow. This ensures that a vessel following a streamline only performs collision avoidance when it gets close to an obstacle; otherwise it will move directly towards the destination, avoiding unnecessary turning and extra fuel consumption. 

The algorithm to generate the stream function is summarized in Algorithm \ref{alg:stream_func}.
\vspace{-1em}
\begin{algorithm}[t]
	\caption{Stream function generation.}
	\label{alg:stream_func}
	\SetKwInput{KwPara}{Parameter}
	\KwIn{$\bm{p}_g$, $\bm{p}_t$, $\bm{p}_i$, $\bm{v}_i$. }
	\KwPara{$r_i$, $l_i$, $C_{\bm{v}_i}$, $N_o$.}
	\KwOut{$\psi_f(\bm{p}_g)$.}  
	\BlankLine
	$Range\_Flag \leftarrow False$;
	
	\For{$\bm{p}_g$ $\in$ $\mathcal{P}_g$}{
		$\psi_f(\bm{p}_g) \leftarrow 0$;
		
		\For{$i \in \mathcal{N}_o$}  
		{  
			\If{$\left| \bm{p}_g-\bm{p}_i \right| \leq l_{i}$}{$Range\_Flag \leftarrow True$; \\
				break;}
		}  
		\eIf{$Range\_Flag$ is $True$}
		{\For{$i \in \mathcal{N}_o$}  
			{  
				\If{$\left| \bm{p}_g-\bm{p}_i \right| \leq l_{i}$}{$\psi_f(\bm{p}_g) \leftarrow \psi_f(\bm{p}_g) + \psi_{i}(\bm{p}_g)+\psi_{\bm{v}_i}(\bm{p}_g)$;}
		} }
		{\For{$i \in \mathcal{N}_o$}  
			{  
				$\psi_f(\bm{p}_g) \leftarrow \psi_f(\bm{p}_g) + \psi_{i}(\bm{p}_g)+\psi_{\bm{v}_i}(\bm{p}_g)$;
		} }
	}
\end{algorithm}

\subsection{Optimal waypoint selection}
Following a constant streamline $\psi_f$, a vessel can eventually reach its destination. However, since the workspace is discretized by a set of grid points, it is difficult to choose the next waypoint $\mathbf{WP}_{k+1}$ with the same value of $\psi_f$ as the current waypoint $\mathbf{WP}_k$. Instead, we choose a next waypoint close to the current waypoint with $\psi_f$ closest to $\psi_f(\mathbf{WP}_k)$. 

\begin{figure}
	\centering
	\subfigure[]{
		\includegraphics[width=0.7\linewidth]{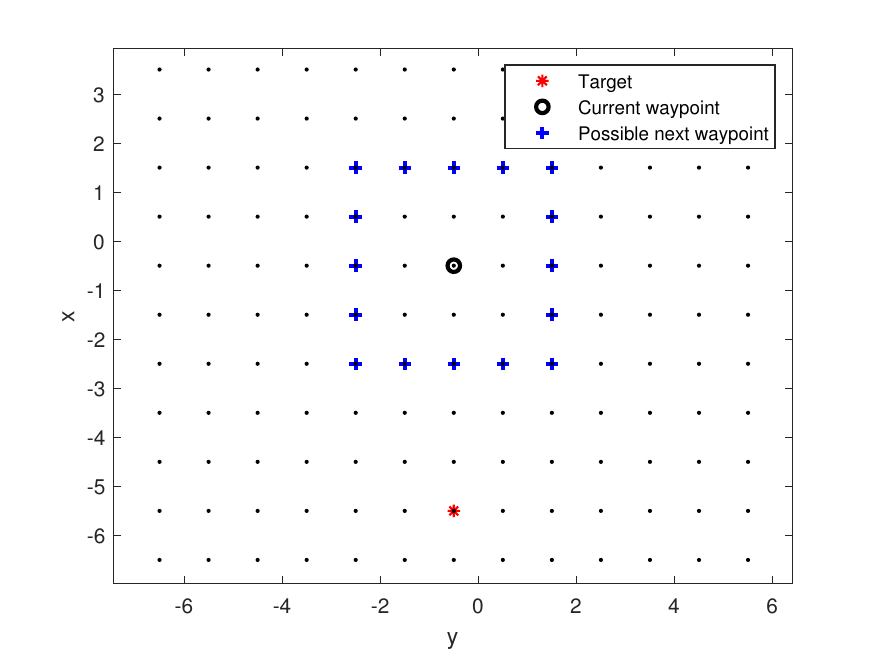}
		\label{Ex_chossing_wpt_a}
	}\\ \vspace{-1em}
	\subfigure[]{
		\includegraphics[width=0.7\linewidth]{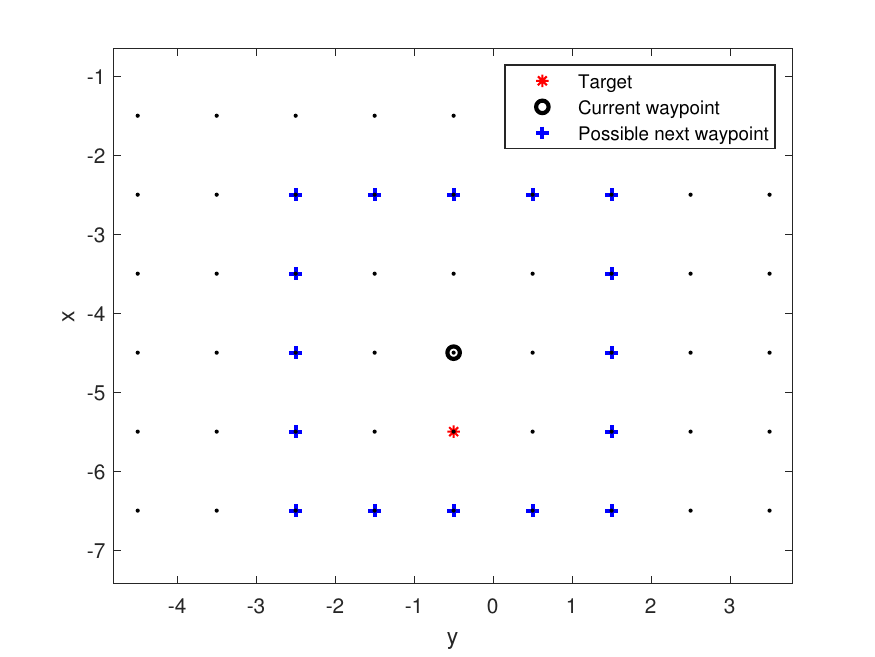}
		\label{Ex_chossing_wpt_b}
	}
	\caption{Illustration of choosing next waypoint with $n_r = 2$. (a) The next waypoint is chosen by solving Eq.~\eqref{SF_waypoint_eq}. (b) The next waypoint is destination.}
	\label{Ex_chossing_wpt} 
\end{figure}

When the positions of the current waypoint $\mathbf{WP}_k=[x_k \ y_k]^\top$ and destination $\bm{p}_{t}=[x_t \ y_t]^\top$ are far apart, illustrated in Figure \ref{Ex_chossing_wpt_a}, the next waypoint is chosen as the solution of the optimization problem,
\begin{subequations}
	\small
	\begin{align}
		\mathbf{WP}_{k+1} & =  \mathop{\arg\min}_{\bm{p}} \   | \psi_f(\bm{p})-\psi_f(\mathbf{WP}_k)| +\gamma \left| \bm{p} -\bm{p}_t \right|  \label{SF_waypoint_eq_a} \\
		\textrm{s.t.} \quad & \bm{p} \in \partial \mathcal{P}_{k} \label{SF_waypoint_eq_b} \\
		& \mathcal{P}_{k} := \left\{ (x, y) \in \mathcal{P}_{g} : \    \max\left\{\frac{|x_k-x|}{d_x}, \frac{|y_k-y|}{d_y}\right\} \leq n_r\right\} \label{SF_waypoint_eq_c}
	\end{align}
	\label{SF_waypoint_eq}%
\end{subequations}
where $\partial$ denotes set boundary, $\gamma$ is a tuning parameter that should be sufficiently small, and the integer $n_r$ is a tuning parameter that reflects the search range of the candidate waypoints.

\begin{figure}[tbh]
	\centering
	\includegraphics[width=0.7\linewidth]{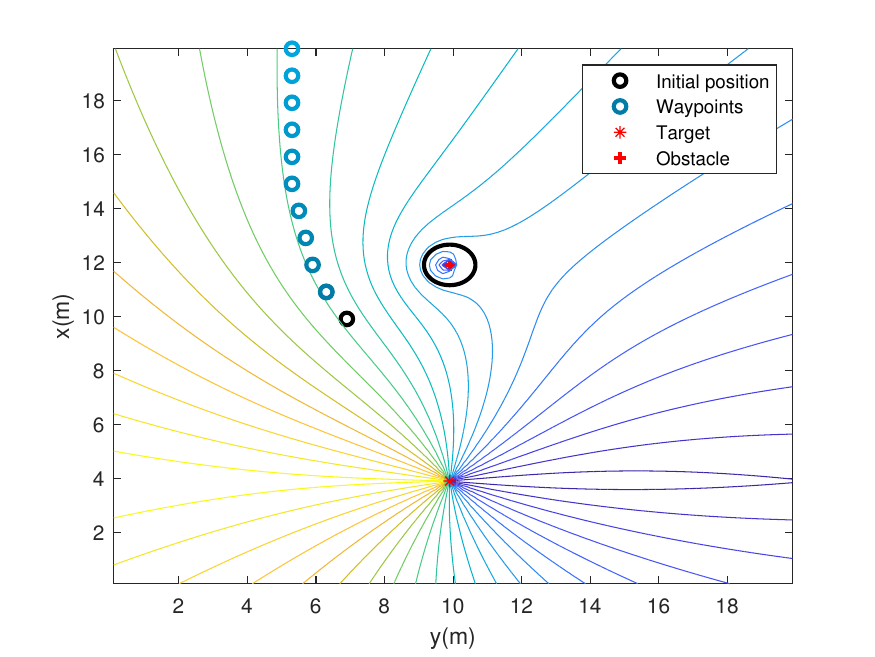}
	\caption{Waypoints that guide the vessel away from the destination if $\gamma = 0$ in Eq.~\eqref{SF_waypoint_eq_a}. The lighter circle represents the waypoint in the future timesteps.}
	\label{Ex_SF_back}
\end{figure}

The vessel following a streamline moves toward the destination, by minimizing the defined cost function \eqref{SF_waypoint_eq_a}. To avoid the vessel being guided away from the destination, the term $\gamma \left| \bm{p}-\bm{p}_t \right|$ is added and penalized, since the streamlines do not give any information about the moving direction. A small $\gamma$ ensures the waypoint close to the destination is always chosen. An example is shown in Figure \ref{Ex_SF_back}. Instead of moving towards the destination, the vessel may be guided away from the destination if we only minimize $| \psi_f(\bm{p})-\psi_f(\mathbf{WP}_k)|$. The candidate waypoints are in the set $\partial \mathcal{P}_{k}$, which are defined in equations~\eqref{eq_grid_point_set} and \eqref{SF_waypoint_eq_c}. An example is shown in Figure \ref{Ex_chossing_wpt_a}. In order to avoid choosing a waypoint very close to the current waypoint, \eqref{SF_waypoint_eq_c} ensures that all candidate waypoints are sufficiently far from the current waypoint according to the $n_r$-multiple of distances $d_x$ and $d_y$.

When the destination is within the rectangular box defined by \eqref{SF_waypoint_eq_c}, shown in Figure \ref{Ex_chossing_wpt_b}, the next waypoint should be the destination, i.e., $\mathbf{WP}_{k+1}:=\bm{p}_{t}, \quad \text{if} \   \bm{p}_t \in  \mathcal{P}_{k}.$

Note that the introduced vortex flows may change the original streamlines near the destination, where the influence is stronger with higher vortex strength and closer distances between the vortices and destination. However, for an alternative streamline, which ends at the destination, can be used to generate the optimal path. With careful parameter tuning, the modified streamlines can be very close to the ones without vortices.

The selection of grid size should satisfy the practical concerns of the traffic density. An impractically large grid interval may result in the vessel being trapped in a local minimum.

The algorithm is summarized in Algorithm \ref{alg:waypoint_selection}, which provides a discrete approximation of the streamlines. While streamlines are guaranteed to provide an obstacle-free path, the discretized approximation is not guaranteed to be obstacle-free. 
The value of $n_r$ influences the update frequency of the optimal waypoint selection. A small number causes unnecessarily high updates where the vessel can be trapped into a local minimum in some extreme situations. On the other hand, the vessel may neglect some dangerous collision if using a too large $n_r$ in dense dynamic traffic. 
For this reason, the value of $n_r$ should be reasonably selected and not too large.

\begin{algorithm}[tbh]
	\caption{Stepwise optimal waypoint selection.}
	\label{alg:waypoint_selection}
	\SetKwInput{KwPara}{Parameter}
	\KwIn{$\mathbf{WP}_k$, $\bm{p}_g$, $\bm{p}_t$, $\bm{p}_i$, $\bm{v}_i$.}
	\KwPara{$d_x$, $d_y$, $n_r$, $\gamma$}
	\KwOut{$\mathbf{WP}_{k+1}$.}  
	\BlankLine
	\eIf{\textnormal{ $\bm{p}_t \not\in \mathcal{P}_{k}$ }}{
		Update stream function using Algorithm \ref{alg:stream_func}; \\
		Determine $\mathbf{WP}_{k+1}$ by Eq. \eqref{SF_waypoint_eq}; \\
	}{$\mathbf{WP}_{k+1} \leftarrow \bm{p}_{t}$;}
\end{algorithm}

\section{Path generation with online stepwise concatenation of path segments}
To implement the proposed guidance model, we integrate it with a path generation module and a path-following controller. In this section, the path generation method using a B\'{e}zier basis is presented, adopted from \cite{Mag2020thesis}.

\subsection{Path generation}
The scenario is this: The vessel is moving on a path segment towards waypoint $\mathbf{WP}_k$ when the next waypoint $\mathbf{WP}_{k+1}$ is determined. The path segment leading to (and through) $\mathbf{WP}_k$ is already fixed, implying that also the path derivatives are defined when approaching $\mathbf{WP}_k$. The problem of path generation is then to construct a feasible path segment from  $\mathbf{WP}_k$ to $\mathbf{WP}_{k+1}$, while ensuring it is sufficiently smooth at $\mathbf{WP}_k$. A path with continuous and bounded third derivatives will be required in the control design \cite{skjetne2005maneuvering}. A desired path segment $k$ in the 2D plane can be represented by
\begin{equation}
	\mathbf{b}_k(\theta): [0,1] \rightarrow \mathbb{R}^2
\end{equation}
where we use $\theta$ as the path parameter.

\subsubsection{The B\'{e}zier curve}
A B\'{e}zier curve of $n$th degree is defined by $n+1$ control points $\mathbf{P}_i$, where $i \in \{0,1,...,n\}$. The first and the last control points define the end points of the curve segment. A B\'{e}zier curve segment $\mathbf{b}(\theta) = [x(\theta) \ y(\theta)]^\top$ parameterized by $\theta\in[0,1]$ can be expressed as a linear combination of the control points \cite{farin2014curves}, i.e.,
\begin{equation}
	\mathbf{b}(\theta) = \mathbf{P}^\top \mathbf{B}^\top \mathbf{a}(\theta),
	\label{bezier_def}
\end{equation}
where
\begin{equation}
	\begin{aligned}
		&\mathbf{a}(\theta) = {\left[\begin{array}{c}
				1   \\
				\theta   \\
				\vdots \\
				\theta^n
			\end{array}\right]},\ \mathbf{B} = {\left[\begin{array}{cccc}
				b_{0,0} & 0 & \dots & 0   \\
				b_{1,0} &  b_{1,1} & \dots & 0   \\
				\vdots & \vdots & \ddots & \vdots\\
				b_{n,0} &  b_{n,1} &  \dots & b_{n,n}
			\end{array}\right]}, \\
		&\mathbf{P} = {\left[\begin{array}{cccc}
				\mathbf{P}_0 \   \mathbf{P}_1 \   \dots  \  \mathbf{P}_n
			\end{array}\right]}^\top = {\left[\begin{array}{cccc}
				x_0 & x_1 &  \dots &  x_n \\
				y_0 & y_1 &  \dots &  y_n 
			\end{array}\right]^\top},
		\label{bezier_def_detail}
	\end{aligned}
\end{equation}
and $b_{i,j}$ is given by \cite{joy2000matrix}
\begin{equation}
	b_{i,j} = (-1)^{i-j}\frac{n!}{(n-i)!}\frac{1}{j!(i-j)!},
\end{equation}
with (!) being the factorial operator.

The derivative of an $n$th order B\'{e}zier curve is still a B\'{e}zier curve but of order $n-1$, given by

\begin{equation}
	\mathbf{b}^\theta(\theta) = \mathbf{P}^\top \mathbf{B}^\top \mathbf{a}^\theta(\theta)
	\label{derivative_Bezier}
\end{equation}

\subsubsection{Optimal control points}
To stepwise generate a feasible path from current to next waypoint, we follow the procedure proposed in \cite{Mag2020thesis}. The path is constrained to lie within a corridor. Within the corridor, the path with minimum length without exceeding the maximum allowed curvature is obtained by an optimization problem. Assuming that we have three waypoints, i.e., the previous $\mathbf{WP}_{k-1}=[x_{k-1}\ y_{k-1}]^\top$, the present $\mathbf{WP}_{k}=[x_{k}\ y_{k}]^\top$, and the next $\mathbf{WP}_{k+1}=[x_{k+1}\ y_{k+1}]^\top$ in the NE frame. The path should have continuous third derivative, i.e., $\mathcal{C}^3$ at the waypoints.  To satisfy this requirement, we choose the degree of a B\'{e}zier curve as $n=7$, which means there are at least 8 control points for each path segment \cite{Mag2020thesis}.

The path segment from $\mathbf{WP}_{k-1}$ to $\mathbf{WP}_{k}$ is generated by a set of control points $\mathbf{P}_{i,k}=[x_{i,k}\ y_{i,k}]^\top$, where $i\in\{0,1,...,7\}$. For $k=1$ (the first path segment), the control points are placed along the straight line from $\mathbf{WP}_{0}$ to $\mathbf{WP}_{1}$ such that 
\begin{subequations}
	\begin{align}
		\mathbf{P}_{i,1}  &= \mathbf{WP}_{0} + \left( \mathbf{WP}_{1}-\mathbf{WP}_{0}\right) \frac{i}{3}\left(\frac{\zeta}{2}\right) , \ i \in \{0,1,2,3 \},  \\
		\mathbf{P}_{i,1}  &= \mathbf{WP}_{1} - \left( \mathbf{WP}_{1}-\mathbf{WP}_{0}\right) \frac{7-i}{3}\left(\frac{\zeta}{2}\right) , \ i \in \{4,5,6,7 \} ,
	\end{align} 
	\label{eq_path_k=0}%
\end{subequations}
where $\frac{\zeta}{2}$ is the radius of a ball embedded within corridors with width $\zeta$ at waypoints $\mathbf{WP}_0$ and $\mathbf{WP}_1$.

\begin{figure}
	\centering
	\includegraphics[width=0.9\linewidth]{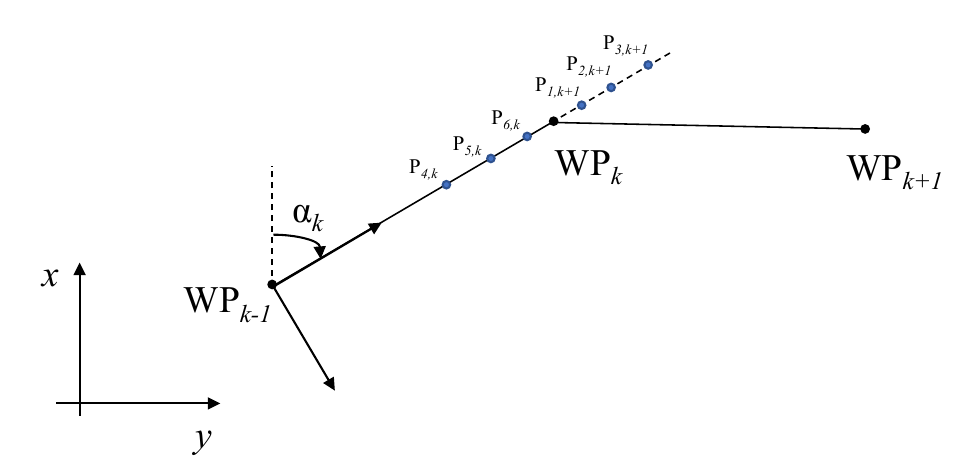}
	\caption{A path-fixed reference frame with origin in $\mathbf{WP}_{k-1}$ rotated by an angle $\alpha_k$ relative to the NE-frame.}
	\label{path_qp_rotation}
\end{figure}

Consider now the path from $\mathbf{WP}_{k-1}$ to $\mathbf{WP}_{k}$ for $k=2,\ 3,\ \hdots$. Then the path segment is obtained by solving an optimization problem. When reaching $\mathbf{WP}_{k}$, the vessel heading should be equal to the path angle $\alpha_{k}=\text{atan2}(y_{k}-y_{k-1},x_{k}-x_{k-1})$. This is achieved by placing $\mathbf{P}_{4,k}$, $\mathbf{P}_{5,k}$, and $\mathbf{P}_{6,k}$ on the straight-line from $\mathbf{WP}_{k-1}$ to $\mathbf{WP}_{k}$, as shown in Figure \ref{path_qp_rotation}. Also, $\mathbf{P}_{1,k+1}$, $\mathbf{P}_{2,k+1}$, and $\mathbf{P}_{3,k+1}$ are uniquely decided by $\mathbf{P}_{7,k}$, $\mathbf{P}_{6,k}$, $\mathbf{P}_{5,k}$, and $\mathbf{P}_{4,k}$, whereas $\mathbf{P}_{0,k+1}$ and $\mathbf{P}_{7,k+1}$ are the end points of the path segment from $\mathbf{WP}_{k}$ to $\mathbf{WP}_{k+1}$. Therefore, $\mathbf{P}_{4,k}$, $\mathbf{P}_{5,k}$, and $\mathbf{P}_{6,k}$ are selected to satisfy
\begin{subequations}
	\begin{align}
		{\left[\begin{array}{ccc}
				1 & 0 & 0   \\
				-2 & 1 & 0   \\
				3 & -3 & 1  
			\end{array}\right]}&
		{\left[\begin{array}{c}
				\mathbf{P}_{1,k+1}   \\
				\mathbf{P}_{2,k+1}   \\
				\mathbf{P}_{3,k+1}   
			\end{array}\right]} \\  &= 
		{\left[\begin{array}{c}
				2\mathbf{P}_{7,k}-\mathbf{P}_{6,k}   \\
				-2\mathbf{P}_{6,k}+\mathbf{P}_{5,k}  \\
				2\mathbf{P}_{7,k}-3\mathbf{P}_{6,k}+3\mathbf{P}_{5,k}-\mathbf{P}_{4,k}  
			\end{array}\right]}, \label{eq_CP_relation} \nonumber\\
		\mathbf{P}_{0,k+1}  &= \mathbf{WP}_{k} , \\
		\mathbf{P}_{7,k+1}  &= \mathbf{WP}_{k+1},
		\label{pragmatic_1}
	\end{align}
\end{subequations}
where the locations of $\mathbf{P}_{i,k+1}$, $i \in \{1,\ 2,\ 3 \}$ must be decided.





There are six decision variables to be optimized, i.e., the $x$- and $y$-axes of the control points $\mathbf{P}_{4,k}$, $\mathbf{P}_{5,k}$, and $\mathbf{P}_{6,k}$ in the NE frame. Since these control points are located on the straight-line path from $\mathbf{WP}_{k-1}$ to $\mathbf{WP}_{k}$, the decision variables can be reduced by expressing $\mathbf{P}_{4,k}$, $\mathbf{P}_{5,k}$, and $\mathbf{P}_{6,k}$ in a path-fixed reference frame, which is rotated by the path angle $\alpha_{k}$ with respect to the NE frame. In the path-fixed reference frame $k$, we denote the $x$-coordinates of all waypoints as $\bm\bar{\chi} = [x_{0,k} \ x_{1,k} \ x_{2,k}\ x_{3,k}\ x_{4,k} \ x_{5,k} \ x_{6,k} \ x_{7,k}]^\top$, where we have omitted index $k$ for brevity. Within $\bm\bar{\chi}$, the decision variables are only the $x$-coordinates $\bm\chi = [x_{4,k} \ x_{5,k} \ x_{6,k}]^\top = [\chi_1 \ \chi_2 \ \chi_3]^\top$, so that $\bm\bar{\chi}=[c^\top,\bm\chi^\top,d^\top]^\top$ with $c$ and $d$ given.

The objective is to minimize the arc length, i.e.,
\begin{equation}
	J = \min_{\bm{\chi}} \int_0^1 \left| \mathbf{b}^\theta(\theta) \right|^2 \mathrm{d}\theta .
	\label{eq_opt}
\end{equation}

Substituting Eq.~\eqref{derivative_Bezier} into \eqref{eq_opt} gives 
\begin{equation}
	\begin{aligned}
		J &= \min_{\bm{\chi}} \int_0^1 \left| \bm\bar{\chi}^\top \mathbf{B}^\top \mathbf{a}^\theta(\theta) \right|^2 \mathrm{d}\theta \\
		&= \min_{\bm{\chi}} \int_0^1  \bm\bar{\chi}^\top \mathbf{B}^\top \mathbf{a}^\theta(\theta)  \mathbf{a}^\theta(\theta)^\top \mathbf{B} \bm\bar{\chi} \mathrm{d}\theta \\  
		&= \min_{\bm{\chi}} \bm\bar{\chi}^\top \mathbf{B}^\top \int_0^1   \mathbf{a}^\theta(\theta) \mathbf{a}^\theta(\theta)^\top   \mathrm{d}\theta \  \mathbf{B} \bm\bar{\chi}.
	\end{aligned}
	\label{eq_opt_middle}
\end{equation}



Note that $\mathbf{B}^\top\int_0^1  \mathbf{a}^\theta(\theta) \mathbf{a}^\theta(\theta)^\top \mathbf{B} \mathrm{d}\theta$ becomes a known constant symmetric positive definite matrix. The objective function contains all $x$-coordinates in $\bm\bar{\chi}$, while only $\bm\chi$ needs to be optimized. By expanding Eq.~\eqref{eq_opt_middle} and neglecting all constant terms, the objective function can, according to \cite{Mag2020thesis}, be simplified to
\begin{equation}
	J = \min_{\bm\chi} \bm\chi^\top \bm{Q} \bm\chi + \bm{q}^\top \bm\chi,
	\label{eq_opt_QP}
\end{equation}
where 
\begin{equation}
	\bm{Q} = \left[\begin{array}{ccc}
		24500    &   -7350     &    980 \\
		-7350    &    2646     &   -441 \\
		980    &    -441     &    98
	\end{array}
	\right],
\end{equation}    
\begin{equation}
	\footnotesize
	\bm{q} = \left[\begin{array}{c}
		8400 x_{0,k} - 41160 x_{1,k} + 82320 x_{2,k} - 85750 x_{3,k} - 70 x_{7,k} \\
		- 1512 x_{0,k} + 8232 x_{1,k} - 18522 x_{2,k} + 22050 x_{3,k} + 42 x_{7,k}  \\
		112 x_{0,k} - 686 x_{1,k} + 1764 x_{2,k} - 2450 x_{3,k} - 14 x_{7,k}
	\end{array}
	\right].
\end{equation}

The constraints of this optimization problem are introduced as follows. First, in Figure \ref{path_qp_rotation}, we notice that $\mathbf{P}_{4,k}$, $\mathbf{P}_{5,k}$, $\mathbf{P}_{6,k}$, $\mathbf{WP}_{k}$, $\mathbf{P}_{1,k+1}$, $\mathbf{P}_{2,k+1}$, and $\mathbf{P}_{3,k+1}$ are arranged in order along the line through $\mathbf{WP}_{k-1}$ and $\mathbf{WP}_{k}$. Similarly, the constraint on $\bm\chi$ is
\begin{equation}
	0 \leq \chi_1 \leq \chi_2 \leq \chi_3  \leq x_{7,k} \leq x_{1,k+1} \leq x_{2,k+1} \leq x_{3,k+1}.
	\label{eq_constraint_arrangement}
\end{equation}

\begin{figure}
	\centering
	\includegraphics[width=0.8\linewidth]{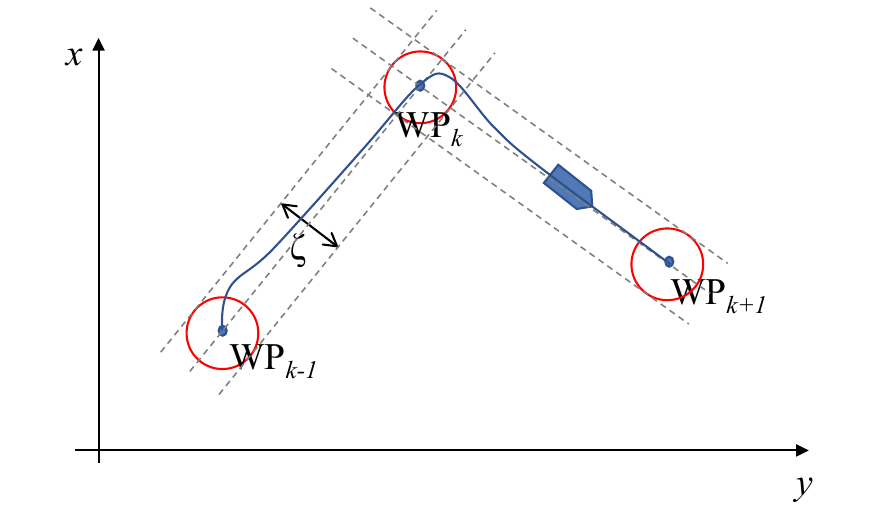}
	\caption{The constraint of corridor on path segment.}
	\label{path_qp_corridor}
\end{figure}

\noindent{Besides, the path segment is designed to stay inside a corridor from $\mathbf{WP}_{k}$ to $\mathbf{WP}_{k+1}$ with a width $\zeta>0$, as illustrated in Figure \ref{path_qp_corridor}. This is achieved by}
\begin{equation}
	x_{i,k+1} \leq  x_{7,k}+\frac{\zeta}{2}, \quad i \in \{ 1,\ 2,\ 3 \} .
	\label{corridor_original}
\end{equation}
where $\frac{\zeta}{2}$ is the radius of a ball embedded within corridors $k$ and $k+1$ at waypoint $\mathbf{WP}_k$.





In addition, the curvature should remain small. Noting that the curvature has its peak right after a given waypoint, the constraints $x_{7,k} \leq x_{1,k+1} \leq x_{2,k+1} \leq x_{3,k+1}$ can be tuned to smooth out sharp turns. This is achieved by introducing a tuning variable $\epsilon$ and forcing $x_{1,k+1}$, $x_{2,k+1}$, and $x_{3,k+1}$ to be greater than $x_{7,k}+\epsilon$ \cite{Mag2020thesis}, that is,
\begin{subequations}
	\begin{align}
		\chi_3 \ &\leq \   x_{7,k}-\epsilon \\
		-\chi_2+4\chi_3 \ &\leq \   3x_{7,k}-\epsilon \\
		\chi_1-6\chi_2+12\chi_3 \ &\leq \   7x_{7,k}-\epsilon,
	\end{align}
	\label{eq_QPconstraint_curvature}%
\end{subequations}
where $\epsilon$ is a small tuning parameter. 

The resulting optimization problem \eqref{eq_opt} with constraints \eqref{eq_constraint_arrangement}, \eqref{corridor_original}, and \eqref{eq_QPconstraint_curvature} is given below. Note that constraints are rewritten in terms of $\bm\chi$.

\begin{equation}
	\begin{array}{rrclcl}
		\displaystyle   \min_{\bm\chi}  & \multicolumn{4}{l}{\bm\chi^\top \bm{Q} \bm\chi + \bm{q}^\top \bm\chi}\\
		\textrm{s.t.} & \mathbf{A}\bm\chi\ &\leq& \ \mathbf{c}\\
		& -\bm\chi \ &\leq& \ \bm0   \\
		& \bm\chi \ &\leq& \ \bm1 x_{7,k}  \\
	\end{array}
	\label{eq_QP}
\end{equation} where $$
\footnotesize{\mathbf{A}^\top = \left[\begin{array}{cccccccccc}
		1 & 0 & 0 & 0 & -1 & 0 & 0 & 1 & 0 & 1 \\
		-1 & 1 & 0 & 1 & 6 & 0 & -1 & -6 & -1 & -5 \\
		0 & -1 & -1 & -4 & -12 & 1 & 4 & 12 & 3 & 8
	\end{array}
	\right]},
$$   
$$
\footnotesize{  \begin{array}{ccccccccccc}
		\mathbf{c}^\top=[ \ 0 & 0 & -x_{7,k+1}+\frac{\zeta}{2} & -3x_{7,k+1}+\frac{\zeta}{2} & \\
		& -7x_{7,k+1}+\frac{\zeta}{2} & x_{7,k+1}-\epsilon  & 3x_{7,k+1}-\epsilon \\
		&  7x_{7,k+1}-\epsilon   & 2x_{7,k+1} & 4x_{7,k+1} \ ].
	\end{array}
}
$$

Using \eqref{eq_path_k=0} for the first segment and solving the optimization problem \eqref{eq_QP} for the later segments, give the set of optimal control points for each path segment $\mathbf{b}_k(\theta)$. Based on $\mathbf{b}_k(\theta)$ obtained using \eqref{bezier_def} for the segment $k$, the desired vessel and path derivatives can be calculated. 

\section{Path-following control}


\subsection{Vessel model}



The vessel is assumed to be fully actuated with motion typically at low speed
so that position and heading are simultaneously controllable. Environmental
disturbances are not considered. The vessel is represented by a 3 degree of
freedom (DOF) maneuvering model \cite{skjetne2020survey}, with motion
restricted to the horizontal plane. The ship heading is referred to as yaw,
while longitudinal and lateral motion is referred to as surge and sway,
respectively.
The control design model is then given by%
\begin{subequations}
	\begin{align} \bm{\dot{p}} &= \mathbf{R}(\psi)\bm{v} \\ {\dot{\psi}} &= r \\ \bm{M}\mathbf{\dot{\bm{\nu}}} + \bm{C}(\bm{\nu})\bm{\nu} + \bm{D}\bm{\nu} &= \bm{\tau}, \end{align} \label{control_design_model}%
\end{subequations}
where $\bm{p}:=[x\ y]^{\top}$ is the ship position in a local North-East
Earth-fixed reference frame with origin $o_{n}$, $\bm{v}:=[u\ v]^{\top}$ is
the surge and sway velocity vector in the body-fixed frame with origin $o_{b}%
$, $\psi$ is the vessel heading/yaw, $r$ is the yaw rate, and
$\bm{\nu}:=[\bm{v}^{\top}\ r]^{\top}$. $\bm{\tau}$ is the thrust load vector,
$\mathbf{R}(\psi)=\left[
\begin{array}
	[c]{cc}%
	\cos\psi & -\sin\psi\\
	\sin\psi & \cos\psi
\end{array}
\right]  $ is the rotation matrix, $\bm{M}$ is the inertia matrix,
$\bm{C}(\bm{\nu})$ is the coriolis and centripetal matrix, and $\bm{D}$ is the
linear damping matrix.

\subsection{The maneuvering control problem}

The path-following control is solved as a maneuvering problem
\cite{skjetne2005maneuvering}. The task of position tracking and heading control is separated in order for the path parameter to only allow feedback from the position state.

\subsubsection{Desired path}
Let $s \in \mathbb{R}$ be an overall continuous path parameter, such that $s=i$ corresponds to $\bm{p} = \mathbf{WP}_i$ for $i=0,1,2,\dots$. Let then 
\begin{align}
	k &=k(s)= \left \lfloor{s}\right \rfloor + 1  \label{eq_k} \\
	\theta &=\theta(s)= s - \left \lfloor{s}\right \rfloor, \label{eq_theta}
\end{align}
where $\left \lfloor{\cdot}\right \rfloor$ is the floor operator.

Given path segment $\mathbf{b}_k(\theta)$, we define the desired path as 
\begin{equation}
	\bm{p}_{d}(s) := \mathbf{b}_k(\theta(s)),
\end{equation}
which by the above procedure is guaranteed to be $\mathcal{C}^3$.



\subsubsection{Control objective}
The control objective for the position and velocity is to satisfy a geometric task and a dynamic task as follows:

\begin{enumerate}
	\item \textbf{Geometric task:} Force the horizontal position $\bm{p}$ to
	converge to and track the desired position $\bm{p}_{d}(s)$, i.e.,
	\begin{equation}
		\lim\limits_{t\rightarrow\infty}|\bm{p}(t)-\bm{p}_{d}(s(t))|=0,
	\end{equation}
	where $\bm{p}_{d}(s)=[x_{d}(s)\ y_{d}(s)]^{\top}$ and $s$ is the path parameter.
	
	\item \textbf{Dynamic task:} Force the path speed $\dot{s}$ to converge to a
	desired velocity $\vartheta_{d}(s)$, i.e.,
	\begin{equation}
		\lim\limits_{t\rightarrow\infty}|\dot{s}(t)-\vartheta_{d}(s(t))|=0.
	\end{equation}
\end{enumerate}

The speed assignment $\vartheta_{d}(s)$ is designed to ensure a constant
desired speed $u_{d}$ along the path. This is achived by
\begin{equation}
	\vartheta_{d}(s):=\frac{u_{d}}{\left\vert p_{d}^{s}(s)\right\vert }=\frac
	{u_{d}}{\sqrt{x_{d}^{s}(s)^{2}+y_{d}^{s}(s)^{2}}}.
\end{equation}

For the heading, the objective is to maintain the direction of the tangent
vector along the path, that is, $\lim\limits_{t\rightarrow\infty}|\psi(t)-\psi
_{d}(s(t))|=0$ where
\begin{equation}
	\psi_{d}(s):=\text{atan}\left(  \frac{y_{d}^{s}(s)}{x_{d}^{s}(s)}\right)  .
	\label{Eq_psid}%
\end{equation}


In the maneuvering control design, the desired parametrized position
$\bm{p}_{d}(s)=[x_{d}(s)\ y_{d}(s)]^{\top}$ is available from the path
generation module, also generating the relevant derivatives with respect to
$s$, that is, $\bm{p}_{d}^{s}(s)=\frac{\partial\bm{p}_{d}(s)}{\partial s}$ and
$\bm{p}_{d}^{s^{2}}(s)=\frac{\partial^{2}\bm{p}_{d}(s)}{\partial s^{2}}$.
Assuming $\dot{s}=\vartheta_{d}(s)$, the heading references are correspondingly
generated from (\ref{Eq_psid}) by
\begin{align}
	\bar{\psi}_{d}(t)  &  :=\psi_{d}(s(t))\\
	\dot{\bar{\psi}}_{d}(t)  &  :=\psi_{d}^{s}(s)\vartheta_{d}(s)\\
	\ddot{\bar{\psi}}_{d}(t)  &  :=\psi_{d}^{s^{2}}(s)\vartheta_{d}(s)^{2}+\psi
	_{d}^{s}(s)\vartheta_{d}^{s}(s)\vartheta_{d}(s).
\end{align}


			\subsection{Backstepping maneuvering control design}
			
			We assume that the full state feedback is available for the control design
			model in \eqref{control_design_model}. A \emph{backstepping} design will
			result in our proposed control law, giving a cascade structure in the error
			states \cite{arcak2001redesign, sorensen2020}. Towards this end, the
			state is transformed by the error signals as $\bm{z}_{p}:=\mathbf{R}%
			(\psi)^{\top}\left(  \bm{p}-\bm{p}_{d}(s)\right)  $, $z_{\psi}:=\psi-\bar
			{\psi}_{d}(t)$, $\bm{z}_{v}:=\bm{v}-\bm{\alpha}_{p}$, $z_{r}:=r-\alpha_{\psi}%
			$, and $\omega:=\dot{s}-\vartheta_{d}(s)$, where
			\begin{align}
				\bm{\alpha}_{p} &  :=-\bm{K}_{p}\bm{z}_{p}+\mathbf{R}(\psi)^{\top}%
				\bm{p}_{d}^{s}(s)\vartheta_{d}(s),\quad\bm{K}_{p}=\bm{K}_{p}^{\top
				}>0\label{Eq_alpha_p}\\
				\alpha_{\psi} &  :=-k_{\psi}z_{\psi}+\dot{\bar{\psi}}_{d}(t),\qquad k_{\psi
				}>0.\label{Eq_alpha_psi}%
			\end{align}
			For the kinematics, this results in%
			\begin{align}
				\dot{\bm{z}}_{p} &  =-\bm{K}_{p}\bm{z}_{p}-r\mathbf{S}\bm{z}_{p}%
				+\bm{z}_{v}-\mathbf{R}(\psi)^{\top}\bm{p}_{d}^{s}(s)\omega\label{Eq_zp_dot}\\
				\dot{z}_{\psi} &  =-k_{\psi}z_{\psi}+z_{r},\label{Eq_zpsi_dot}%
			\end{align}
			where $\mathbf{S}={\left[
				\begin{array}
					[c]{cc}%
					0 & -1\\
					1 & 0
				\end{array}
				\right]  =-}\mathbf{S}^{\top}$. Let the first step control Lyapunov function be
			$V_{1}(\bm{z}_{p},z_{\psi})=V_{p}(\bm{z}_{p})+V_{\psi}(z_{\psi}):=\frac{1}%
			{2}\bm{z}_{p}^{\top}\bm{z}_{p}+\frac{1}{2}z_{\psi}^{2}$, giving
			\begin{equation}
				\dot{V}_{1}=-\bm{z}_{p}^{\top}\bm{K}_{p}\bm{z}_{p}-k_{\psi}z_{\psi}%
				^{2}+\bm{z}_{p}^{\top}\bm{z}_{v}+z_{\psi}z_{r}-\bm{z}_{p}^{\top}%
				\mathbf{R}(\psi)^{\top}\bm{p}_{d}^{s}(s)\omega.\label{Eq_V1dot}%
			\end{equation}
			While the cross-terms with $\bm{z}_{v}$ and $z_{r}$ are handled in the kinetic
			control design next, we proceed first by assigning the update law for $\omega$
			to render the last term of (\ref{Eq_V1dot}) nonpositive. Several choices are
			possible \cite{rskj2020}. Noting that
			$V_{p}^{s}=\frac{\partial V_{p}}{\partial s}=-\bm{z}_{p}^{\top}\mathbf{R}%
			(\psi)^{\top}\bm{p}_{d}^{s}(s)$ and $\mathbf{R}(\psi)\bm{z}_{p}%
			=\bm{p}-\bm{p}_{d}(s)$,\ we propose
			\begin{equation}
				\omega=-\frac{\mu}{\left\vert \bm{p}_{d}^{s}(s)\right\vert +\varepsilon}%
				V_{p}^{s}=\mu\frac{\bm{p}_{d}^{s}(s)^{\top}}{\left\vert \bm{p}_{d}%
					^{s}(s)\right\vert +\varepsilon}\left(  \bm{p}-\bm{p}_{d}(s)\right)
				,\ \mu\geq0,\label{eq_omega}
			\end{equation}
			referred to as the \emph{unit-tangent gradient update law}, where
			$0<\varepsilon\ll1$ is a regularization constant. This update law ensures for
			$\mu$ large to keep $V_{p}(\bm{z}_{p}(p,\psi,s))$ minimized w.r.t. $s$, that
			is, by maintaining the location $\bm{p}_{d}(s)$ along the path where the inner
			product between the (approximate unit) tangent vector $\frac{\bm{p}_{d}%
				^{s}(s)^{\top}}{\left\vert \bm{p}_{d}^{s}(s)\right\vert +\varepsilon}$ and
			error vector $\bm{p}-\bm{p}_{d}(s)$ is approximately zero.
			
			We proceed with the kinetic design, defining $\bm{z}_{\nu}%
			:=\bm{\nu}-\bm{\alpha}$ and $\bm{\alpha}:=[\bm{\alpha}_{p}^{\top}%
			\ \alpha_{\psi}]^{\top}$, where $\dot{\bm{\alpha}}$ is available through
			differentiation of (\ref{Eq_alpha_p})--(\ref{Eq_alpha_psi}). The control law is
			assigned as
			\begin{equation}
				\bm{\tau}=-\bm{K}_{\nu}\bm{z}_{\nu}-\bm{C}(\mathbf{\bm{\nu}}%
				)\bm{\nu}+\bm{D}\bm{\alpha}+\bm{M}\dot{\bm{\alpha}}, \label{eq_tau}%
			\end{equation}
			so that the resulting closed-loop system becomes (\ref{Eq_zp_dot}%
			)-(\ref{Eq_zpsi_dot}) and
			\begin{equation}
				\bm{M}\dot{\bm{z}}_{\nu}=-\left(  \bm{K}_{\nu}+\bm{D}\right)  \bm{z}_{\nu},
				\label{Eq_znu_dot}%
			\end{equation}
			where $\bm{K}_{\nu}$ is a diagonal positive definite matrix.
			
			\begin{theorem}
				Under the assumption that $\bm{p}_{d}(s)$ and its path derivatives are
				bounded, the noncompact set
				\begin{equation}
					\mathcal{A}:=\left\{  (\bm{z}_{p},z_{\psi},\bm{z}_{\nu},s,t):\ [\bm{z}_{p}%
					^{\top},z_{\psi},\bm{z}_{\nu}^{\top}]=0\right\}
				\end{equation}
				is uniformly globally exponentially stable (UGES) for the\ closed-loop system
				(\ref{Eq_zp_dot})--(\ref{Eq_zpsi_dot}), (\ref{Eq_znu_dot}), and
				\begin{equation}
					\dot{s}=\vartheta_{d}(s)+\mu\frac{\bm{p}_{d}^{s}(s)^{\top}}{\left\vert
						\bm{p}_{d}^{s}(s)\right\vert +\varepsilon}\left(  \bm{p}-\bm{p}_{d}(s)\right)
					.\label{Eq_s_dot}%
				\end{equation}
				
				\begin{proof}
					The closed-loop error dynamics is a cascade where the UGES kinetic error
					system (\ref{Eq_znu_dot}) is connected to the UGES kinematic error systems
					(\ref{Eq_zp_dot})--(\ref{Eq_zpsi_dot}) and (\ref{eq_omega}). From standard cascade theory
					\cite[Proposition 2.3]{Lamnabhi2005},
					this ensures that the cascaded error system is UGES on the maximal interval of
					existence. This implies that the states $(\bm{z}_{p},z_{\psi},\bm{z}_{\nu})$
					are bounded on the maximal interval of existence. By also boundedness of the
					path signals, this ensures that the right-hand side of (\ref{Eq_s_dot}) is
					bounded, thus precluding finite escape time. UGES of $\mathcal{A}$ 
					follows from the results in \cite{skjetne2005maneuvering}. 
				\end{proof}
			\end{theorem}
			
			The selection of control gains should be tuned as a practical tradeoff between convergence speed, capacity of the propeller system, and power consumption.

\section{Case studies}
\subsection{Simulation overview}
The pseudocode for the proposed system is summarized in Algorithm \ref{alg:GNC}. In the pseudocode, we first initialize the parameters and environments (Line 1). Then, as long as the ship is not at the destination, the code enters the \textit{while} loop, which includes optimal waypoints selection, path generation, and maneuvering control (Lines 2-18). Lines 2-11 ensures that once the ship reaches a waypoint $\mathbf{WP}_k$, it finds the optimal next waypoint $\mathbf{WP}_{k+1}$ (Lines 3-4) and calculates the control points for path generation (Lines 5-9). Note that we only find the next waypoint, instead of all subsequent waypoints leading to the destination. Finally, we apply maneuvering control law and update the ship states (Lines 12-17).

\begin{algorithm}[t]
	\caption{The pseudocode for the proposed system.}
	\label{alg:GNC}
	\BlankLine
	Initialize $k = 0$, $k_{s} = 1$, $t = 0$, $s(0) = 0$, $\bm{p}$, $\bm{p}_t$, $\bm{p}_i$, $\bm{v}_i$, $\mathbf{WP}_k = \bm{p}$; \\
	\While{\textnormal{$\left| \bm{p}_t - \bm{p} \right| \geq \delta$}}{
		\If{ $k_{s} > k$}
		{
			Find $\mathbf{WP}_{k+1}$ using Algorithm \ref{alg:waypoint_selection}; \\
			\eIf{$k = 0$}
			{Calculate control points using Eq. \eqref{eq_path_k=0};}
			{Calculate optimal control points by solving problem \eqref{eq_QP};}
			$k \leftarrow k_{s}$;
		}
		Calculate $\dot{s}=\omega+\vartheta_{d}(s)$ and $s = \int_0^t \dot{s}(\tau) \mathrm{d}\tau$  ;  \\		
		Calculate $k_{s}$ by Eq. \eqref{eq_k}; \\ 
		Calculate $\theta$ by Eq. \eqref{eq_theta}; \\ 
		Calculate $\bm{p}_d$ based on control points and $\theta$ using Eq. \eqref{bezier_def}; \\ 
		Calculate $\bm\tau$ using Eq. \eqref{eq_tau}; \\
		Apply $\bm\tau$ and update states in the control design model \eqref{control_design_model}; \\
	}
	
\end{algorithm}

Simulations are performed on the control design model of the CyberShip Enterprise I (CSEI). Six simulation cases are designed to test the proposed methods, including overtaking, head-on, crossing, and complex situations. The first two simulations show the results when all vessels are guided by the stream-function algorithm presented in this paper. In the last four simulations, all obstacles have constant speeds following straight-line trajectories, without COLREGs compliance. The results show that the proposed method is able to guide a marine vessel to its destination with both collision avoidance and COLREGs compliance in dynamic and complex situations. Therefore, we demonstrate in two simulation cases that the proposed method has the following features:
\begin{enumerate}
	\item COLREGs Compliance: Maneuvering own ship and target ships where all these apply our COLREG-compliant algorithm.
	\item Obstacle Avoidance: Anti-collision maneuvers of own ship in environment with moving obstacles that all have constant velocity vectors (not obeying COLREGs).
\end{enumerate}

The data for obstacles are listed in Table \ref{Obstacle_para_1} and Table \ref{Obstacle_para_2}, where each row of the matrices represents the parameters in $x$- and $y$-directions, respectively, for an obstacle as indexed in figures \ref{Result_intelligent_1} - \ref{Result_v_complex2}. The matrices for CSEI are given in \cite{CSE1}. The vehicle dynamics is taken into account by the entire system during the path generation and control design.







The $20$ m $\times 20$ m workspace is represented by the discrete grid with $d_x = d_y = 0.2$ m. We choose $\gamma=0.2$, $\delta=0.01$, $n_r=5$, and $r_i=l_i=1.5$. For the stream function, we choose $\psi_{\mathbf{d,l}} = \frac{\pi}{4}$ and $\psi_{\mathbf{d,u}} = \frac{3\pi}{4}$. To generate a feasible path, we choose $\zeta = 0.5$ and $\epsilon = 0.005$. The parameters for the maneuvering controller is chosen as $\bm{K}_p=\text{diag}([20,20])$, $k_{\psi} = 40$, $\bm{K}_{\nu}=\text{diag}([20,20,20])$, $u_d = 0.2$, $\varepsilon = 0.01$, and $\mu = 10^{-4}$. The ship has initial heading $\frac{\pi}{2}$.

The tuning process for waypoint selection is very simple due to the small number of tuned parameters. The main parameters include $d_x=d_y$, $n_r$, $\gamma$, and $C_{\bm{v}_i}$. Small $d_x$ and $d_y$ give a fine grid that can capture the workspace information accurately. A too fine grid, on the other hand, can be computationally expensive. Hence, $d_x$ and $d_y$ should be tuned with a trade-off between accuracy and efficiency. Given $d_x$ and $d_y$, $n_r$ represents the frequency of waypoint selection; smaller $n_r$ means deciding a new waypoint more frequently. $\gamma$ should be a small number, which prevents the vessel from being guided away due to grid discretization. The tuning of $C_{\bm{v}_i}$ depends on $\bm{v}_i$, since $C_{\bm{v}_i}\left|\bm{v}_i\right|$ represents the strength of the vortex flow.

			
			
			

\begin{table}
	\centering
	\resizebox{\linewidth}{!}{  
		\begin{threeparttable}
			\caption{Obstacle parameters in COLREG-compliant maneuvers.}
			\label{Obstacle_para_1}
			\begin{tabular}{cccc}
				\toprule
				Scenario &  \begin{tabular}{@{}c@{}}Initial position \\ (m)\end{tabular} & \begin{tabular}{@{}c@{}}Velocity vector\\ (m/s) \end{tabular}   &  $C_{\bm{v}_i}\left|\bm{v}_i\right|$   \cr
				\midrule
				\begin{tabular}{@{}c@{}}Head-on and \\ overtaking\end{tabular}  & $\begin{bmatrix}11.9& 9.9\\5.9& 9.9\end{bmatrix}$ & $\begin{bmatrix}3.9& 9.9\\17.9& 9.9\end{bmatrix}$ & $\begin{bmatrix}0.1\\0.1\end{bmatrix}$ \cr\cr
				Crossing situation  & $\begin{bmatrix}15.9& 15.9\end{bmatrix}$ & $\begin{bmatrix}3.9&3.9\end{bmatrix}$ & $\begin{bmatrix}0.1\end{bmatrix}$ \cr
				\bottomrule
			\end{tabular}
		\end{threeparttable}
	}
\end{table}

\begin{table}
	\centering
	\resizebox{\linewidth}{!}{  
		\begin{threeparttable}
			\caption{Obstacle parameters in anti-collision maneuvers.}
			\label{Obstacle_para_2}
			\begin{tabular}{cccc}
				\toprule
				Scenario &  \begin{tabular}{@{}c@{}}Initial position \\ (m)\end{tabular} & \begin{tabular}{@{}c@{}}Velocity vector\\ (m/s) \end{tabular}   &  $C_{\bm{v}_i}\left|\bm{v}_i\right|$   \cr
				\midrule
				
				\begin{tabular}{@{}c@{}}Head-on \\ situation\end{tabular} & $\begin{bmatrix}5.9& 9.9\\3.9& 7.9\\1.9& 11.9\end{bmatrix}$ & $\begin{bmatrix}0.04& 0\\0.04& 0\\0.04& 0\end{bmatrix}$ & $\begin{bmatrix}0.05\\0.1\\0.05\end{bmatrix}$ \cr\cr
				
				\begin{tabular}{@{}c@{}}Crossing \\ situation\end{tabular} & $\begin{bmatrix}14.9& 11.9\\11.9& 5.9\\8.9& 16.9\\5.9& 1.9\end{bmatrix}$ & $\begin{bmatrix}0& -0.04\\0& 0.04\\0& -0.04\\0& 0.04\end{bmatrix}$ & $\begin{bmatrix} 0.05\\0.08\\0.1\\0.1 \end{bmatrix}$ \cr\cr
				
				\begin{tabular}{@{}c@{}}Complex \\ situation 1\end{tabular} & $\begin{bmatrix}14.9& 9.9\\11.9& 11.9\\9.9& 13.9\\5.9& 17.9\\ 4.9& 5.9\end{bmatrix}$ & $\begin{bmatrix}0.04& 0.04\\ 0.024& -0.04\\0.008& -0.056\\0& -0.024\\0.024& 0.024 \end{bmatrix}$ & $\begin{bmatrix}0.05\\
					0.05\\0.05\\0.1\\0.1\end{bmatrix}$ \cr\cr
				
				\begin{tabular}{@{}c@{}}Complex \\ situation 2\end{tabular} & $\begin{bmatrix}12.9& 9.9\\7.9& 13.9\\9.9& 15.9\\3.9& 1.9\\ 1.9& 4.9\\3.9& 17.9\end{bmatrix}$ & $\begin{bmatrix}0.04& 0\\ 0.016& -0.04\\0.024& -0.04\\0.008& 0.04\\0.04& 0.04\\-0.04& 0 \end{bmatrix}$  & $\begin{bmatrix}0.06\\ 0.06\\0.1\\0.1\\0.1\\0.1 \end{bmatrix}$  \cr
				\bottomrule
			\end{tabular}
		\end{threeparttable}
	}
\end{table}

\subsection{Case 1: Simulation results of COLREGs-compliance}

\begin{figure}[htb!]
	\centering
	\includegraphics[width=0.8\linewidth]{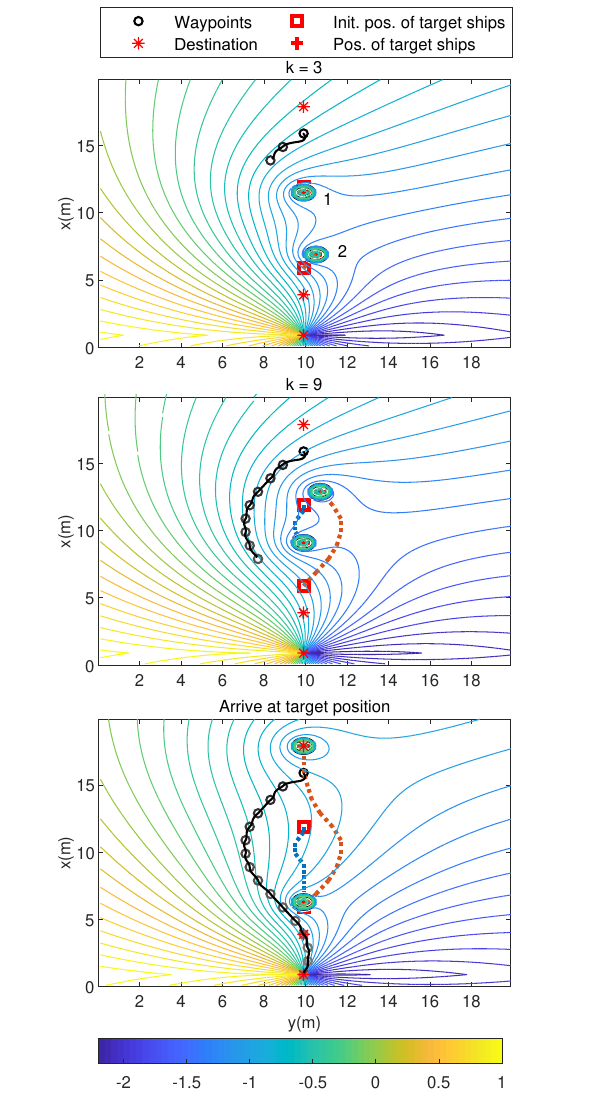}
	\caption{Simulation result of the COLREG-compliant head-on/overtaking situation. The own ship is moving from North to South. The streamlines are plotted from the perspective of the own ship.}
	\label{Result_intelligent_1}
\end{figure}

\begin{figure}[htb!]
	\centering
	\includegraphics[width=0.8\linewidth]{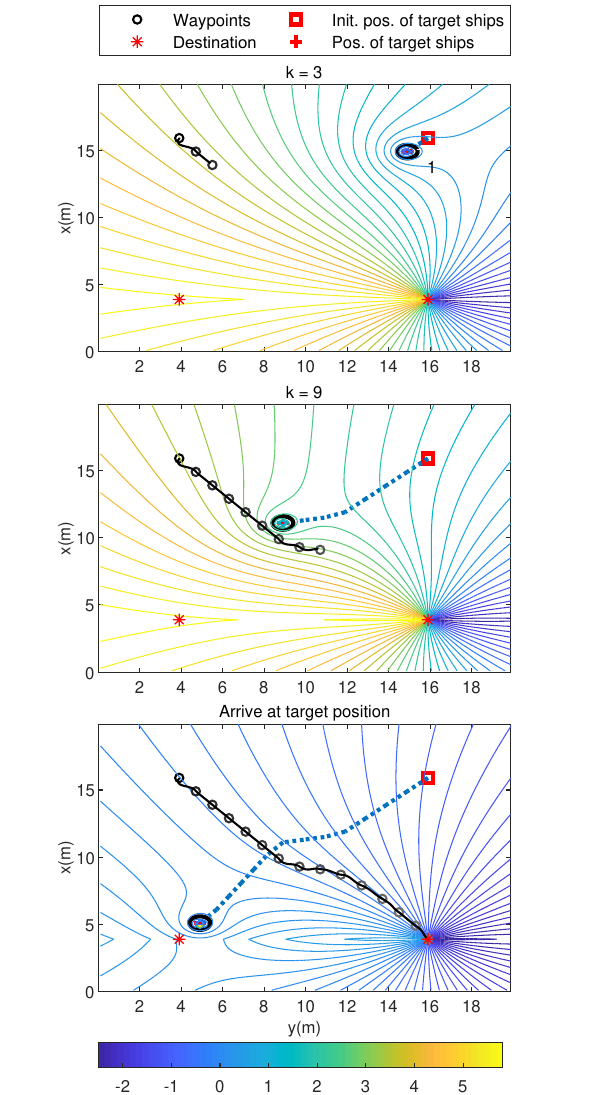}
	\caption{Simulation result of the COLREG-compliant crossing situation. The own ship is moving from NorthWest to SouthEast. The streamlines are plotted from the perspective of the own ship.}
	\label{Result_intelligent_2}
\end{figure}

The simulation results are presented in figures \ref{Result_intelligent_1} and \ref{Result_intelligent_2}. The vessel (own ship) and the obstacles (target ships) all use the proposed stream function method with $S_i(\psi_{\mathbf{d}}) = 1$ to choose waypoints. This setting renders all vessels to have COLREGs-compliant behaviors. For simplicity, we have implemented the target ships to have straight-line trajectories from one waypoint to the next.

\subsubsection{Head-on and overtaking situation}
The initial position of the OS is located at $[15.9 \  9.9]^\top$, and its destination position is $\bm{p}_t=[0.9 \  9.9]^\top$. TS1 intends to move towards south slowly, while TS2 moves towards north, creating a head-on situation. With $S_i(\mathbf{d}) = 1$, they both move to starboard to avoid collision. During its voyage to the destination, the OS moves to starboard and pass the two TS.

\subsubsection{Crossing situation}
The initial position of the OS is located at $[15.9 \  3.9]^\top$, with its destination at $\bm{p}_t=[3.9 \  15.9]^\top$. In this situation, the OS and the TS have symmetric positions and destinations. Since the OS crosses from the starboard side of the TS, the OS has right of way. At $k=9$, we can see that the TS gives way and the OS crosses in front of the TS, which comply with COLREGs Rule 15. We do note that, even though the OS has right of way, it does make a slight starboard maneuver. This behavior is conservative, and it may be mitigated by, for instance, incorporating additional logic that omits the crossing TS when generating the stream function guidance model. 

\begin{figure}[htb!]
	\centering
	\includegraphics[width=0.8\linewidth]{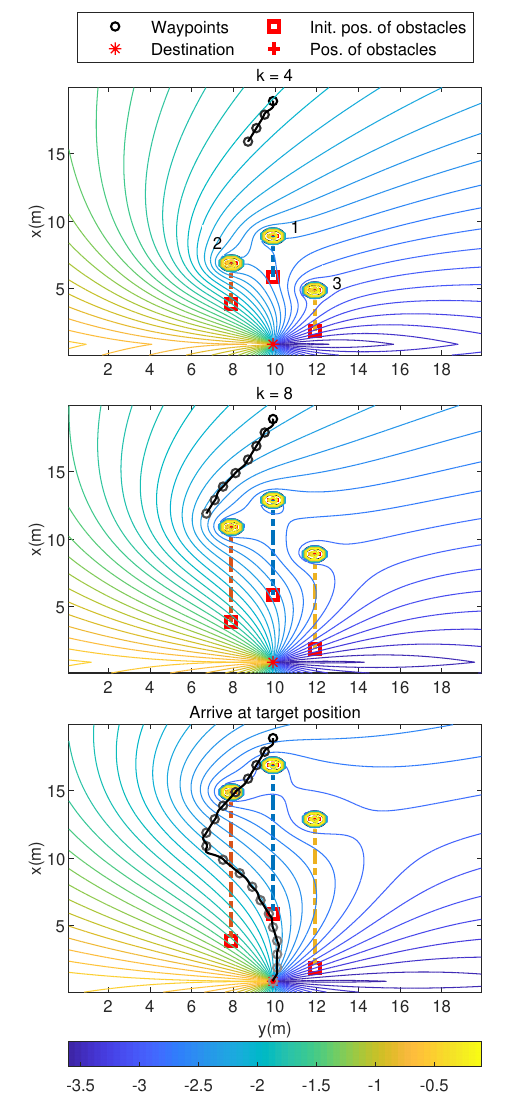}
	\caption{Obstacle avoidance simulation result of the head-on situation. The own ship is moving from North to South.}
	\label{Result_v_headon}
\end{figure}

\begin{figure}[htb!]
	\centering
	\includegraphics[width=0.75\linewidth]{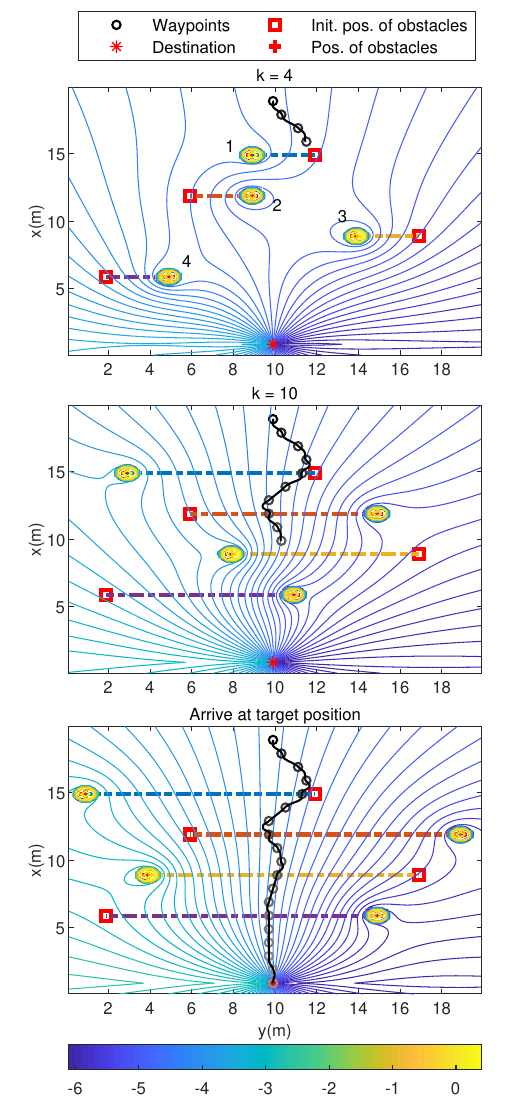}
	\caption{Obstacle avoidance simulation result of the crossing situation. The own ship is moving from North to South.}
	\label{Result_v_crossing}
\end{figure}

\begin{figure}[t]
	\centering
	\includegraphics[width=0.75\linewidth]{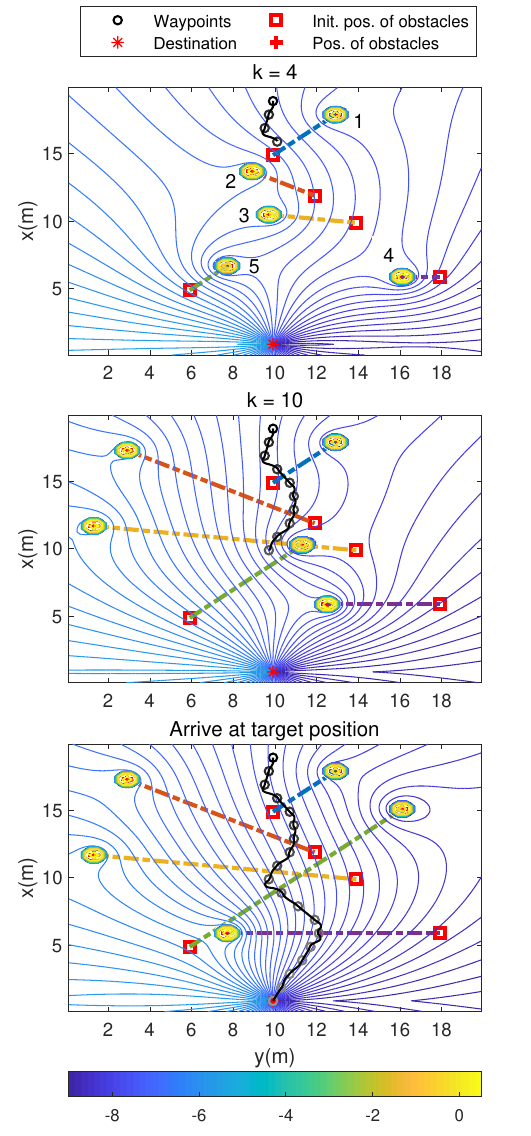}
	\caption{Obstacle avoidance simulation result of the complex situation 1. The own ship is moving from North to South.}
	\label{Result_v_complex1}
\end{figure}

\begin{figure}[htb!]
	\centering
	\includegraphics[width=0.75\linewidth]{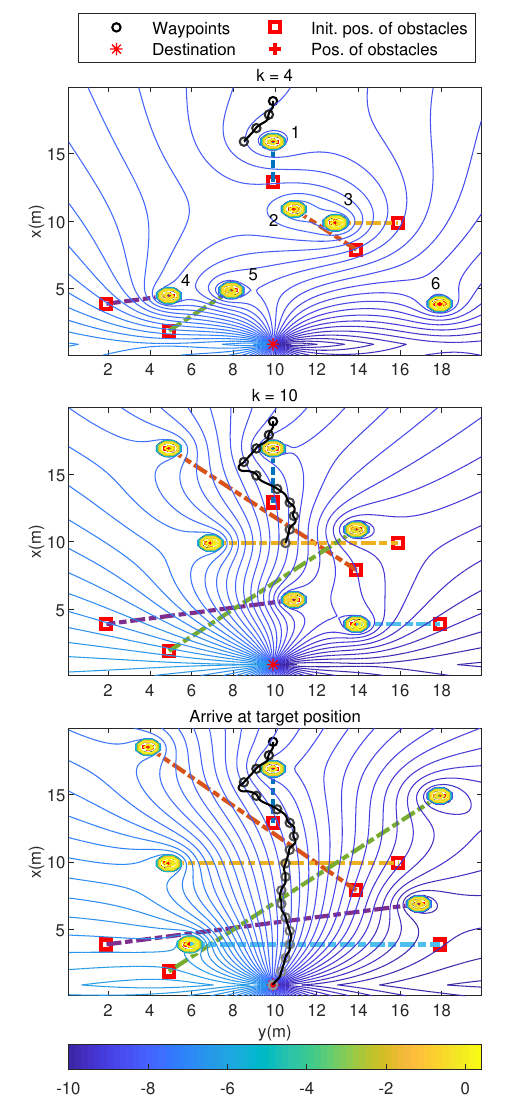}
	\caption{Obstacle avoidance simulation result of the complex situation 2. The own ship is moving from North to South.}
	\label{Result_v_complex2}
\end{figure}

\subsection{Case 2: Simulation results of obstacle avoidance}
Results of the proposed method are presented in figures \ref{Result_v_headon} - \ref{Result_v_complex2}. The initial position of our vessel is located at $[18.9 \  9.9]^\top$, and its destination is $\bm{p}_t=[0.9 \  9.9]^\top$. In these simulations, all obstacles move with constant velocities without COLREGs compliance. 

To highlight the performance of the proposed method, we compare the results with the method in \cite{waydo2003vehicle}; see more details in the Appendix.
Adding doublets affects local streamlines around the obstacles and renders shorter paths. The results in figures \ref{SF_benchmark_5}, \ref{SF_benchmark_8}, and \ref{SF_benchmark_9} indicate that this may cause closer encounters and a higher risk.
We recognize this, however, as an alternative method to our proposed vortex flows.

\subsubsection{Head-on situation}
The result of the head-on situation is illustrated in Figure \ref{Result_v_headon}. There are three obstacles moving in the northern direction, creating a head-on situation. Similar to the overtaking situation, the flow in the stream function method has three counterclockwise vortex flows, and this makes the vessel move to starboard and passing the obstacles safely on its port side.

\subsubsection{Crossing situation}
The result of the crossing situation is shown in Figure \ref{Result_v_crossing}. There are four obstacles, with obstacles 1 and 3 moving in the westward direction and obstacles 2 and 4 in the eastward direction. By definition, $S_i(\psi_{\mathbf{d}}) = -1$ (clockwise vortex flow) for obstacles 1 and 3, and $S_i(\psi_{\mathbf{d}}) = 1$ (counterclockwise vortex flow) for obstacles 2 and 4. In such a flow field, it can be seen that at $k=4$ the vessel moves behind Obstacle 1. By following the streamline, the vessel also avoids crossing in front of obstacles 2 and 3, which gives an S-shaped trajectory at $k=10$. After that, the vessel is able to move almost directly towards the destination with no risk of collision as Obstacle 4 is maintaining its course.

\subsubsection{Complex situation 1}
The result of the first complex situation is presented in Figure \ref{Result_v_complex1}. There are five obstacles moving in different directions, creating a complex situation. We have $S_i(\psi_{\mathbf{d}}) = -1$ for obstacles 2, 3, and 4, and $S_i(\psi_{\mathbf{d}}) = 1$ for obstacles 1 and 5. At $k=4$, we can see that the vessel first moves to starboard to avoid Obstacle 1, then changes its course to port to avoid collision with obstacles 2 and 3. At $k=10$, the vessel has obstacle 5 on its port side to ensure safety. Thereafter, the vessel behavior to avoid Obstacle 4 is similar to the crossing situation presented in Figure \ref{Result_v_crossing}.

\subsubsection{Complex situation 2}
The result of the second complex situation is shown in Figure \ref{Result_v_complex2}. There are six obstacles moving in different directions. We have $S_i(\psi_{\mathbf{d}}) = -1$ for obstacles 2, 3, and 6, and $S_i(\psi_{\mathbf{d}}) = 1$ for obstacles 1, 4, and 5. In the beginning, there is a head-on situation and the vessel moves to starboard, similar to the result in Figure \ref{Result_v_headon}. At $k=4$, it starts to change its course to port to avoid collision with obstacle 2. At $k=10$, it can be seen that the vessel follows the streamline between obstacles 3 and 5, which is beneficial to both make progress towards the destination and avoid collision with obstacles 3 and 5. The vessel behavior thereafter to avoid obstacles 4 and 6 is also similar to the crossing situation.

From the simulation results, we notice that a drawback for the stream function method is that, after adding vortex flows, not all streamlines end perfectly at the destination; see figures \ref{Result_v_complex1} and \ref{Result_v_complex2}. A potential solution is to only add vortex components within a user-defined range of the obstacles. In practice, however, the proposed method is acceptable due to the insignificant deviation from the destination, and since the vessel will typically enter another control mode in its continuing operation when approaching the destination.


\section{Conclusions}
This paper presented a method to accomplish autonomous guidance and stepwise path planning with anti-collision and COLREGs compliance. The stream function, augmented with vortex flows, is adopted to generate waypoints. The method was integrated with a path generation algorithm using B\'{e}zier curves, which formulates a quadratic programming problem to minimize the path length between two waypoints. The dynamics of the marine vessel is taken into account by adding constraints on path curvature. A backstepping-based maneuvering control design, resulting in a cascade structure in the error states, was performed to achieve path following. The system has been demonstrated through simulations, where six scenarios were presented, including overtaking, head-on, crossing, and complex situations.

In future, more COLREGs rules and other regulations can be used to assess the proposed methods. By incorporating more rules, increasingly complex situations can be studied. The system can also be tested under model uncertainties and environmental disturbances. It would be interesting to add a motion prediction module to estimate obstacles' trajectories, and assess the robustness of the proposed methods. Besides, inter-vessel communication during COLREGs situations should be incorporated into the design. In addition, the complex marine environment and shipboard power/propulsion should be considered in future studies by including the uncertain current speed, wave spectrum, and thruster configuration and capacity.

\bibliographystyle{IEEEtran}
\bibliography{Reference.bib} 

\addtolength{\textheight}{-12cm}   


\section*{APPENDIX}

\begin{figure}[htb!]
	\centering
		\subfigure[]{
		\includegraphics[width=0.5\linewidth]{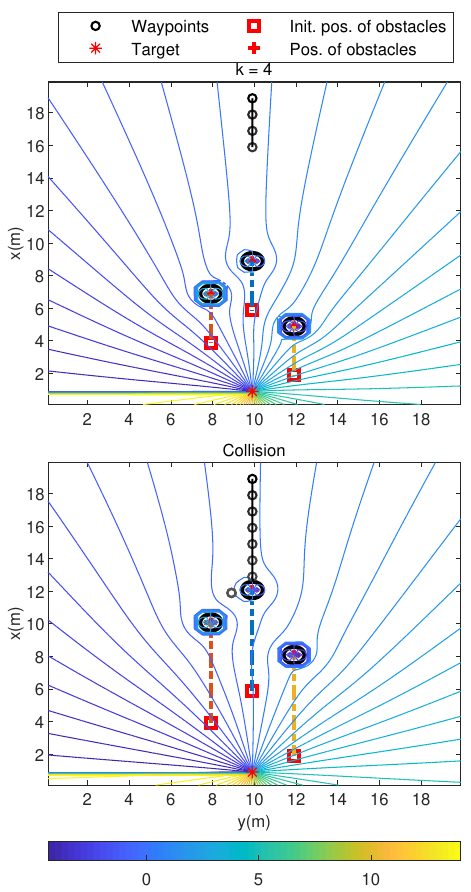}
		\label{SF_benchmark_5}
	} \\
	\subfigure[]{
		\includegraphics[width=0.5\linewidth]{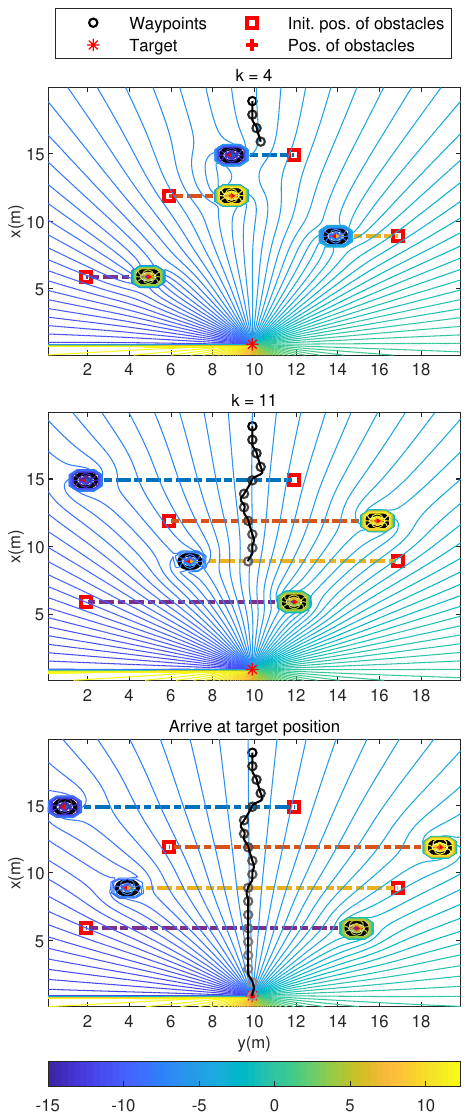}
		\label{SF_benchmark_7}
	} \\
	\caption{Obstacle avoidance simulation result of the head-on and crossing situation using the method in \cite{waydo2003vehicle}.}
	\label{SF_benchmark_1} 
\end{figure}

\begin{figure}[htb!]
	\centering
	\subfigure[]{
		\includegraphics[width=0.5\linewidth]{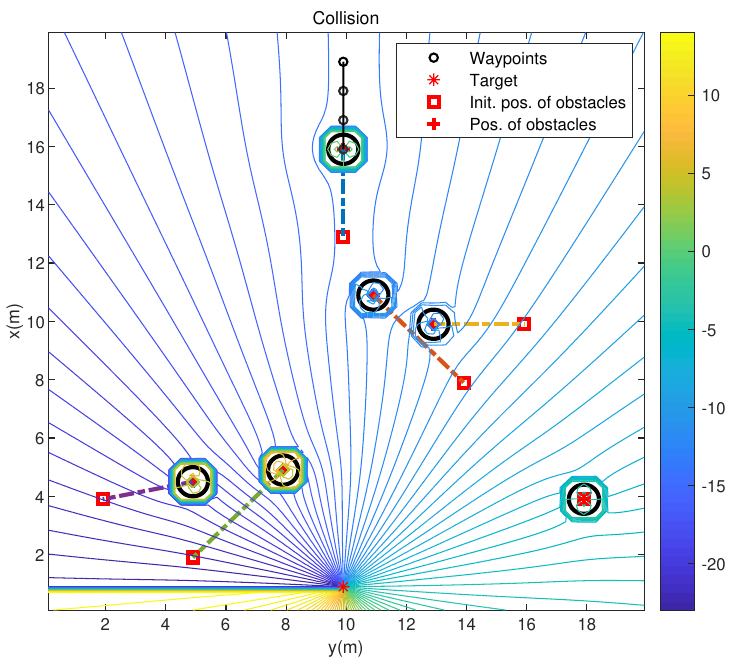}
		\label{SF_benchmark_8}
	} \\ 
	\subfigure[]{
		\includegraphics[width=0.5\linewidth]{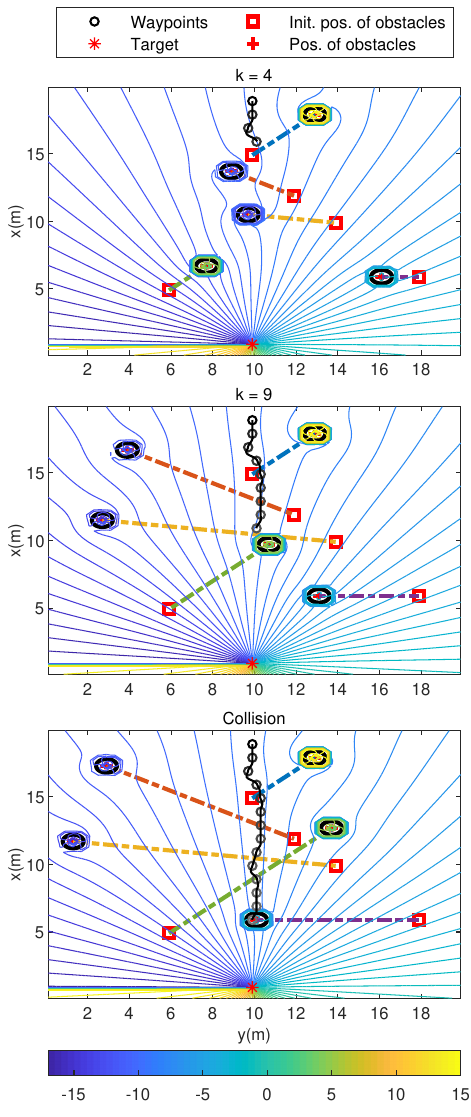}
		\label{SF_benchmark_9}
	}
	\caption{Obstacle avoidance simulation result of the complex situations using the method in \cite{waydo2003vehicle}.}
	\label{SF_benchmark_2} 
\end{figure}

Simulation results of obstacle avoidance using the method in \cite{waydo2003vehicle} are illustrated in figures \ref{SF_benchmark_1} - \ref{SF_benchmark_2}. The complex potential for a moving obstacle with velocity $\bm{v} = [x_{v}  \  y_{v}]^\top$ is \cite{waydo2003vehicle}

\begin{equation}
	\omega(\mathcal{Z}) = \omega_f(\mathcal{Z}) - y_{v} \left(\frac{\mathrm{a}^2}{\mathcal{Z}-\mathrm{b}}+\Bar{\mathrm{b}}\right) - \mathrm{i} x_{v} \left(\frac{\mathrm{a}^2}{\mathcal{Z}-\mathrm{b}}+\Bar{\mathrm{b}}\right),
\end{equation} 
where $\omega_f(\mathcal{Z})$ is the complex potential obtain by applying the Circle theorem \eqref{eq_Circletheorem}. It is proved that the streamlines become tangent to the obstacle boundary \cite{waydo2003vehicle}. The additional terms $- y_{v} \left(\frac{\mathrm{a}^2}{\mathcal{Z}-\mathrm{b}}+\Bar{\mathrm{b}}\right) - \mathrm{i} x_{v} \left(\frac{\mathrm{a}^2}{\mathcal{Z}-\mathrm{b}}+\Bar{\mathrm{b}}\right)$ is the potential flow of a doublet which describes how the vehicle should pass around the moving obstacle to avoid collision. The strength of this doublet is decided by the velocity of the moving obstacle. To extend to the setting with multiple moving obstacles, addition and thresholding is applied with $l_{i} = \frac{3}{2} r_{i}$.

%
%
%

\end{document}